\documentstyle[12pt,epsf]{article}
\newcommand{\be}{\begin{equation}}
\newcommand{\ee}{\end{equation}}
\newcommand{\bea}{\begin{eqnarray}}
\newcommand{\eea}{\end{eqnarray}}
\newcommand{\brr}{\begin{array}}
\newcommand{\err}{\end{array}}
\newcommand{\bc}{\begin{center}}
\newcommand{\ec}{\end{center}}
\newcommand{\nn}{\nonumber}

\newcommand{\ms}{m_s^{\overline{MS}}(\mu)}
\newcommand{\as}{\alpha_{ s}}

\newcommand{\ep}{\epsilon}
\newcommand{\epp}{\epsilon^{\prime}}
\newcommand{\epse}{\epsilon^{\prime}/\epsilon}

\newcommand{\GF}{\frac{G_{ F}}{\sqrt 2}}

\newcommand{\gammazeros}{\hat{\gamma}^{(0)T}_{ s}}
\newcommand{\gammazeroe}{\hat{\gamma}^{(0)T}_{ e}}
\newcommand{\gammaones}{\hat{\gamma}^{(1)T}_{ s}}
\newcommand{\gammaonee}{\hat{\gamma}^{(1)T}_{ e}}

\newcommand{\DSone}{\Delta S\!=\!1}
\newcommand{\DStwo}{\Delta S\!=\!2}

\newcommand{\Kzero}{K^{ 0}}
\newcommand{\Kbzero}{{\bar K}^{ 0}}

\newcommand{\md}{m_{ d}}

\newcommand{\J}{\hat{J}}
\newcommand{\Ke}{\hat{ K}}
\newcommand{\U}{\hat{U}}

\newcommand{\W}{\hat{     W}}

\newcommand{\PP}{\hat{ P}}

\newcommand{\Gmulup}{\gamma^{\mu}_{ L}}

\newcommand{\alphas}{\alpha_{ s}}
\newcommand{\alphae}{\alpha_{ e}}
\begin{document}
\pagestyle{empty}
\begin{flushright}
CERN-TH.7514/94 \\
ROME prep. 94/1024
\end{flushright}
\centerline{\bf{An Upgraded Analysis of $\epp/\ep$
  at the Next-to-Leading Order}}
\vskip 1cm
\centerline{\bf{ M. Ciuchini$^{a}$,
E. Franco$^b$, G. Martinelli$^{c,*}$, L. Reina$^d$ and
L. Silvestrini$^e$}}
\centerline{$^a$ INFN, Sezione Sanit\`a, V.le Regina Elena 299,
00161 Roma, Italy. }
\centerline{$^b$ Dip. di Fisica,
Universit\`a degli Studi di Roma ``La Sapienza" and}
\centerline{INFN, Sezione di Roma, P.le A. Moro 2, 00185 Roma, Italy. }
\centerline{$^c$ Theory Division, CERN, 1211 Geneva 23, Switzerland.}
\centerline{$^d$ Brookhaven National Laboratory, Physics Department, Upton,
NY 11973}
\centerline{$^e$ Dip. di Fisica, Univ. di Roma ``Tor Vergata''
and INFN, Sezione di Roma II,}
\centerline{Via della Ricerca Scientifica 1, I-00133 Rome, Italy.}
\begin{abstract}
An upgraded analysis of $\ep$, $x_d$ and  $\epp/\ep$,
using the latest determinations of the relevant experimental and
theoretical parameters, is presented. Using the recent determination
of the top quark mass, $m_t=(174 \pm 17)$ GeV, our best
estimate  is
$\epp/\ep= 3.1  \pm 2.5 $, which lies in the range given by E731.
We describe  our determination
of $\epp/\ep$ and make a comparison with other similar studies.
A detailed discussion of the matching of the full theory to
the effective Hamiltonian, written in terms of lattice operators,
is also given.
\end{abstract}
\vskip 3.0cm
\begin{flushleft}
CERN-TH.7514/94 \\
January 1995
\vskip 1 cm
$^*$On leave of absence
from  Dip. di Fisica,
Universit\`a degli Studi di Roma ``La Sapienza".
\end{flushleft}

\vfill\eject
\pagestyle{empty}\clearpage
\setcounter{page}{1}
\pagestyle{plain}
\newpage
\pagestyle{plain} \setcounter{page}{1}
\section{Introduction}
\protect\label{sec:intro}
The understanding of mixing and  $CP$ violation in hadronic systems
constitutes one  of the crucial tests of the
Standard Model. In the last few years considerable
theoretical and experimental effort has been invested in this sector.
\par
On the experimental side,
more accurate measurements of the
mixing angles are now  available and the mass of the top quark, recently
discovered, is constrained within tight limits \cite{top}. Still, in spite
of very accurate measurements, the experimental results for the $CP$
violating parameter $\epse$
are far from conclusive \cite{na31,e731}.
\par  On the theoretical side,
the complete next-to-leading expressions of the relevant effective
$\Delta S=1$, $\Delta S=2$, $\Delta B=1$
 and $\Delta B=2$ Ha\-mil\-to\-nians have been
computed \cite{alta}--\cite{noi}, thus reducing the theoretical
 uncertainties\footnote{ Indeed only the top contribution to
the $\Delta S=2$ Hamiltonian is fully known,
at the next-to-leading order.
To our knowledge,
the other contributions have been only
partially computed \cite{hn}.}. Moreover, there is now
increasing theoretical evidence that the value of the
pseudoscalar $B$-meson decay constant is large, $f_B \sim 200$ MeV, and that
the $B^0$--$\bar B^0$ mixing parameter $B_B$ is quite close to one.
Still, the evaluation of other  hadronic matrix elements is subject to large
uncertainties, which are particularly severe for $\epse$, where
important cancellations of different contributions occur,
for large values of the top mass. A real
improvement in the calculation of the hadronic matrix elements, from
lattice simulations, or with other non-perturbative
techniques, is still missing.
In spite of this, we believe that, given the
experimental and theoretical novelties,  it is time to present an
upgraded analysis
of mixing and $CP$ violation for $K$- and $B$-mesons,
along the lines followed in refs. \cite{reina,ciuc1}.
We will also
have the opportunity of presenting our results for the Wilson
coefficients of the operators appearing in the
$\Delta S=1$ Hamiltonian in different regularization/renormalization
schemes, which may be useful for further phenomenological analyses by
other authors. We also intend to present and
  clarify several issues related to the
matching conditions of  effective Hamiltonians to the full
theory, to the use of the lattice regularization, to the value of the
mass of the strange quark and to  quark thresholds in the evolution
of the Wilson coefficients.
Particular emphasis is devoted to a realistic evaluation of the
uncertainties in our predictions.
\par
The results given here contain several improvements with
 respect to our previous work on the same subject \cite{ciuc1}.
The main differences are the following:
\begin{itemize}
\item Upgraded  values of the experimental parameters entering in
the phenomenological analysis,
such as the $B$-meson lifetime $\tau_B$,
the $B^0_d$--$\bar B^0_d$ mixing parameter $x_d$,
the CKM matrix elements, ($\vert V_{cb} \vert$,
$\vert V_{ub}\vert/\vert V_{cb}\vert $),  etc., have been used.
\item The value of the strange quark mass $m_s$ has been taken from
lattice calculations, thus making
a more consistent use of  lattice results for
the $B$-parameters of the relevant penguin operators \cite{ms}.
\item The relevant formulae, necessary to express
the effective  Hamiltonian, derived in the full theory,   are given
in terms of lattice  operators.
\item Assumptions and relations which have been used to
evaluate the $B$-parameters of the operators entering  the calculation,
in particular those that have not been computed either on the lattice, or with
other non-perturbative techniques, are critically reviewed.
\item Uncertainties, coming from the matching between lattice and
con\-ti\-nuum operators, are discussed, together with those
 due to the choice of the
renormalization scale, $\Lambda_{QCD}$, the mass of the top quark, etc.
We also discuss different determinations of $\epp/\ep$ originating from the
choice of
different continuum  regularization schemes.
\item All the results are presented with an estimate of the corresponding
errors. These errors come from the limited precision of the measured
quantities,
e.g. $\tau_B$, and from theoretical uncertainties, e.g. the
value of hadronic matrix elements.
\end{itemize}
The phenomenological results of this study
have been partially reported elsewhere
 \cite{viet,glasgow}. In this paper we
put particular emphasis on some theoretical aspects and subtleties
related to the construction of the effective Hamiltonian. The reader
will also find a practical tool to combine the Wilson coefficients
with his own preferred results for the operator matrix elements, sections
\ref{sec:coe}--\ref{sec:ri}.
\par
The plan of the paper is the following.
We  summarize in  section \ref{sec:results}
the main results obtained  from a combined analysis of
$\ep$, $x_d$ and $\epse$  together with our estimate of the
errors. All the  details of the analysis will be given after this
section.
In section \ref{sec:hs},
we give the relevant formulae of the effective Hamiltonians, which control
$\ep$, $B^0$--$\bar B^0$ transitions and $\epp/\ep$,
and define the relevant operators.
In section \ref{sec:coe}, a summary  of the calculation of the
Wilson coefficients of the operators, at the next-to-leading order, is given.
The matching of the continuum theory with the
 lattice is addressed in section \ref{sec:latcon}. The formulae
presented in this section  have been used to derive
the relation between the mass of the quarks on the lattice and in the
continuum. In section \ref{sec:ri} we give details on
the Regularization Independent ($RI$) renormalization prescription,
together with the numerical values
of the Wilson coefficients in different renormalization schemes.
In section \ref{sec:bpar}, the values and errors of the
$B$-parameters, used in this analysis, are given
and compared with those used in ref. \cite{burasepe}. A
discussion of the assumptions and relations among the $B$-parameters is also
presented in the same section.
The main uncertainties of the phenomenological analysis
are reviewed  in sec. \ref{sec:error}.
\section{Results}
\protect\label{sec:results}
In this section, the main results of the present study are summarized.
These results have been obtained by varying the experimental quantities,
e.g. the value of the top mass $m_t$, $\tau_B$,  etc., and
the theoretical parameters, e.g. the kaon $B$-parameter $B_K$ or
the strange quark mass $\ms$, according to their errors.
Values and errors of the input quantities, used in the following, are reported
in tables \ref{tab:val}-- \ref{tab:bpar}.
For the experimental quantities, we assume a Gaussian distribution and,
for the theoretical ones, a flat distribution (with a width of  $2\,\sigma$).
The only exception is $\ms$, taken from quenched lattice $QCD$ calculations,
for which we have assumed a Gaussian distribution, according to the results
of ref. \cite{ms}.
\par
\begin{figure}   
    \centering
    \epsfxsize=1.08\textwidth
    \leavevmode\epsffile{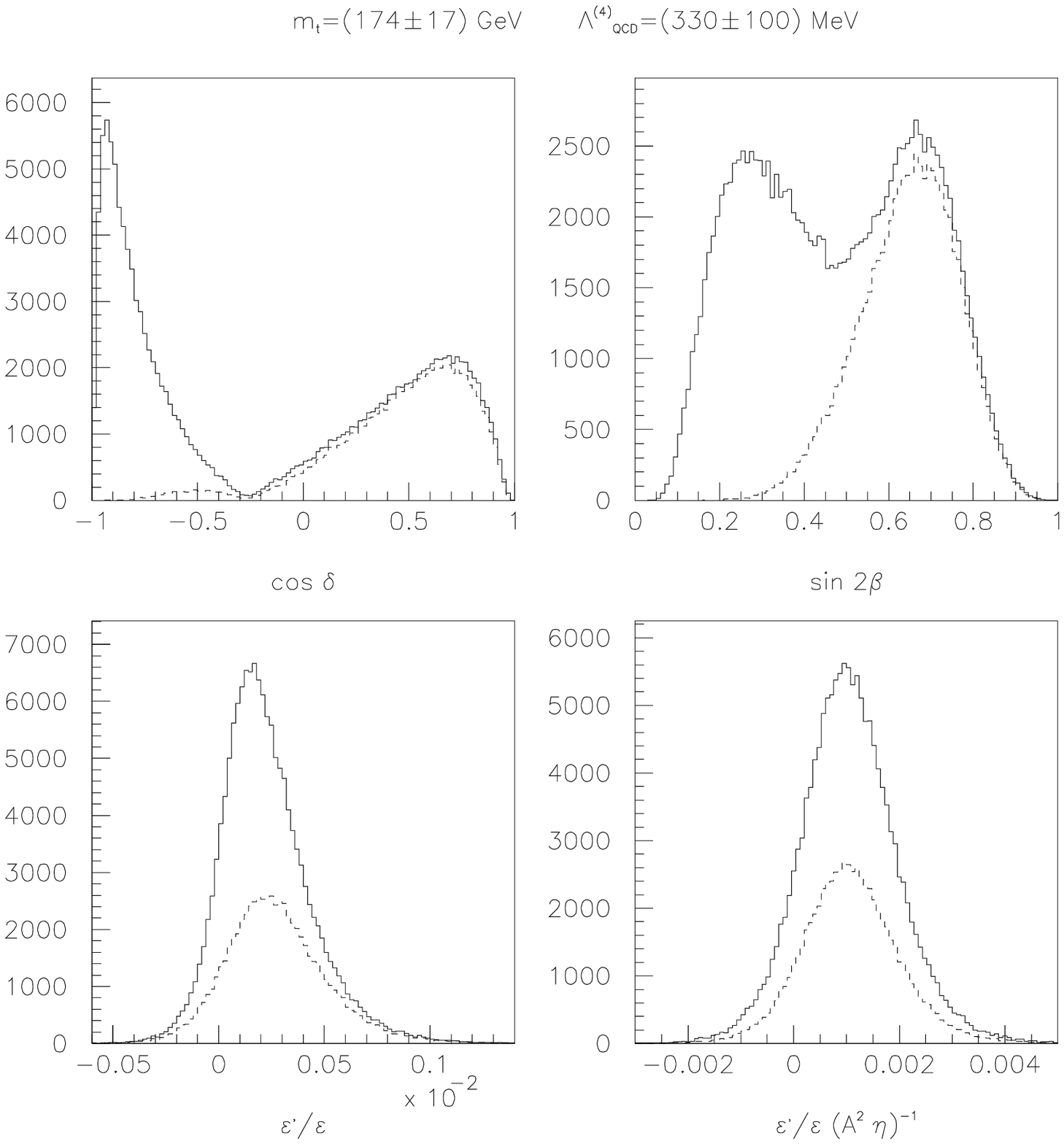}
       \caption[]{\it{ Distributions of values for $\cos \delta$,
$\sin 2 \beta$, $\epse$ and $\epse (A^2\eta)^{-1}$,
for $m_t=(174 \pm 17)$ GeV,
and using the values of the parameters given in tables \ref{tab:val}--
\ref{tab:bpar}. The solid histograms are obtained without using
the information coming from  $B_d$--$\bar B_d$ mixing, the dashed ones
after imposing that $160$ MeV $\le f_B B_B^{1/2}\le 240$ MeV. }}
    \protect\protect\label{fig:cd}
\end{figure}

Each theoretical
prediction ($\cos \delta$, $\sin 2\beta$, $\epse$,  etc.),
depends on several input parameters, which fluctuate independently.
In many cases, this produces a pseudo-Gaussian distribution of values,
see for example $\epse$ in fig. \ref{fig:cd}. From the width of the
pseudo-Gaussian, we estimate the errors of our predictions.
Non-Gaussian distributions may also occur, for example
when a single input parameter dominates the result.
In these cases, we still estimate the error from the
variance of the distribution.
\begin{table}
\begin{center}
\begin{tabular}{|c|c|}
\hline\hline
Parameters & Values \\ \hline
$m_t$ & $(174 \pm 17)$ GeV \\
$m_s$(2 GeV) & ($128\pm 18$) MeV \\
$\Lambda_{QCD}^{n_f=5}$ & ($230 \pm 80$) MeV \\
$V_{cb}=A\lambda^2$ & $0.040\pm 0.006$ \\
$\vert V_{ub}/V_{cb}\vert=\lambda\sigma$ & $0.080\pm 0.015$ \\
$\tau_B$ & $(1.49\pm 0.12)\times 10^{-12}$ sec \\
$x_d$ & $0.685\pm 0.076$ \\
$(f_B B_B^{1/2})_{th}$ & $(200\pm 40)$ MeV \\
$\Omega_{IB}$ & $0.25\pm 0.10$ \\ \hline\hline
\end{tabular}
\caption[]{ \it{Values and errors of the parameters used in the numerical
analysis.}}
\protect\label{tab:val}
\end{center}
\end{table}

\begin{table}
\begin{center}
\begin{tabular}{|c|c|}
\hline\hline
Constants & Values \\ \hline

$G_F$ & $1.16634\times 10^{-5}\mbox{GeV}^{-2}$ \\
$m_c$ & 1.5 GeV \\
$m_b$ & 4.5 GeV\\
$M_W$ & 80.2 GeV \\
$M_{\pi}$ & 140 MeV \\
$M_K$ & 490 MeV \\
$M_B$ & 5.278 GeV \\
$\Delta M_K$ & $3.521\times 10^{-12}$ MeV \\
$f_{\pi}$ & 132 MeV \\
$f_K$ & 160 MeV \\
$\lambda=\sin\theta_c$ & 0.221 \\
$\ep_{exp}$ & $2.268\times 10^{-3}$ \\
$\mbox{Re}A_0$ & $2.7\times 10^{-7}$ GeV \\
$\omega$ & 0.045 \\
$\mu$ & $2$ GeV\\ \hline\hline
\end{tabular}
\caption[]{ \it{Constants used in the numerical analysis. }}
\protect\label{tab:const}
\end{center}
\end{table}
\begin{table}
\begin{center}
\begin{tabular}{|c|c|c|c|c|c|c|}\hline\hline
$B_K$ & $B_9^{(3/2)}$ &  $B_{1-2}^{ c}$ &
 $B_{3,4}$ &
$B_{5,6}$ & $B_{7-8-9}^{(1/2)}$ & $B_{7-8}^{(3/2)}$\\
\hline
$0.75\pm 0.15$ & $0.62\pm 0.10$  &  $0 - 0.15^{(*)}$ & $1 -  6^{(*)}$ &
$1.0\pm 0.2$ & $1^{(*)}$ & $1.0\pm0.2$
\\ \hline\hline
\end{tabular}
\caption[]{{\it Values of the $B$-parameters, for operators renormalized at the
scale $\mu=2$ GeV. The only exception is $B_K$, which is the renormalization
group-invariant $B$-parameter, defined in ref. \cite{slom}; $B_9^{3/2}(\mu)$
has been taken equal to $B_K(\mu)$, at any renormalization scale
$\mu$.
In the table, we give  $B_9^{3/2}(\mu=2$  GeV$)$.
Entries with a $^{ (*)}$
are educated guesses; the others are taken from lattice $QCD$ calculations
\cite{shar3}--\cite{shar4}. See section \ref{sec:bpar} for more details.}}
\protect\label{tab:bpar}
\end{center}
\end{table}

\begin{figure}   
    \centering
    \epsfxsize=1.08\textwidth
    \leavevmode\epsffile{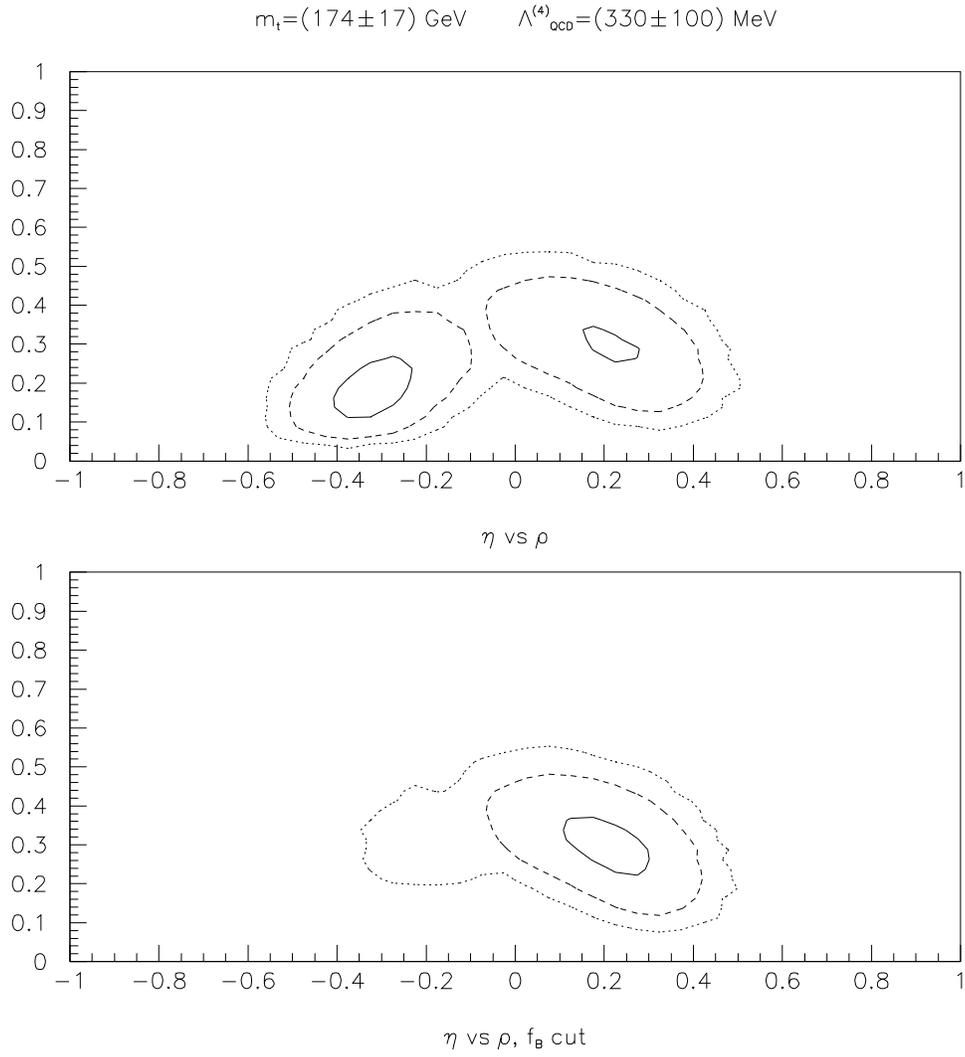}
       \caption[]{\it{ Contour plots in the $\rho$--$\eta$ plane.
The solid, dashed and dotted contours contain
$5 \%$, $68 \%$ and $95 \%$ of the generated events respectively.
The contours are given by excluding or including the
$f_B$-cut. Similar results can be found in refs. \cite{schubert,alig}.}}
\protect\label{fig:rhoeta}
\end{figure}

Using the numbers given in the tables and the formulae given in the
forthcoming section, we have obtained the following
results:
\par a)  The comparison of the experimental value of $\ep$ with its theoretical
prediction will in general correspond to two possible solutions for the
value of $\cos \delta$, see for example ref. \cite{reina}. Because of the
errors of the input parameters, the two solutions appear as a distribution
of values, with a twin-peak shape. The distribution
for $\cos \delta$ is given in fig. \ref{fig:cd}. As already noticed in
refs. \cite{reina,ciuc1} and \cite{schubert,alig},
a large value of $f_B$, combined with a large $m_t$, favours $\cos \delta >
0$. When the condition $ 160 $ MeV $\le f_B B_B^{1/2}
\le 240$ MeV is imposed ($f_B$-cut),
most of the negative solutions  disappear,
giving the dashed histogram of fig. \ref{fig:cd}, from which we estimate
\be \cos \delta= 0.47 \pm 0.32\,\, . \protect\label{eq:cd} \ee
\par b)  $\cos \delta$ is correlated to the values of the Wolfenstein
parameters $\rho$ and $\eta$. In fig. \ref{fig:rhoeta}, the contour plot
in the $\rho$--$\eta$ plane is given, with and without the $f_B$-cut.
%
\par c) The value of $\sin 2 \beta$ depends on
 $\cos \delta$
\be \sin 2 \beta =  \frac{2 \sigma \sin \delta (1-\sigma \cos \delta)
}{ 1 + \sigma^2 -2 \sigma \cos \delta}. \ee

\begin{figure}[t]   
\begin{center} \setlength{\unitlength}{1truecm}
\begin{picture}(6.0,7.0)
\put(-6.0,-7.0){\includegraphics{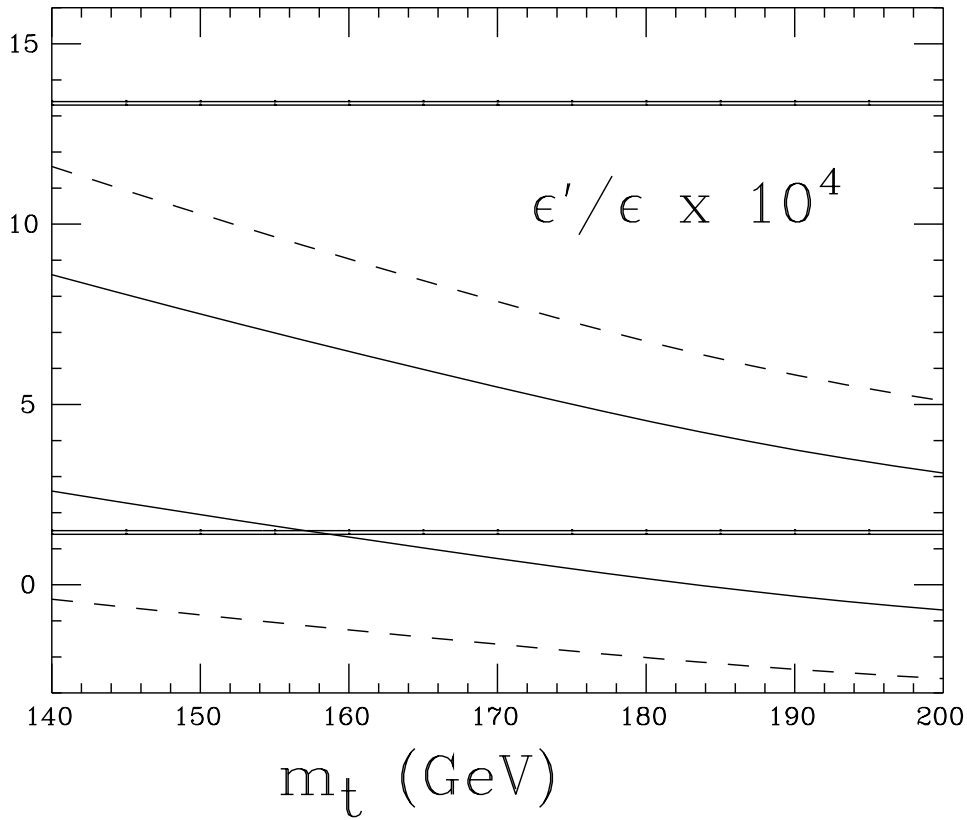}}
\end{picture} \end{center}
\vskip 1.5 cm
    \caption[]{\it{ $\epse$  as a function of $m_t$,
obtained by applying the $f_B$-cut. The zones delimited by
the solid and dashed curves represent the allowed
regions containing respectively $68\%$ and $95\%$ of the generated events.
The region between the two double lines is the experimental result
coming from E731 \cite{e731}, $\epse=(7.4 \pm 5.9) \times 10^{-4}$.
The range from NA31 \cite{na31}, $\epse=(23.0\pm 6.5)\times 10^{-4}$ is
not given in the figure.}}
\protect\label{fig:mtd}
\end{figure}

 The distribution of $\sin 2 \beta$ is shown in
fig. \ref{fig:cd}, without (solid)  and with (dashed) the $f_B$-cut.
When the $f_B$-cut is imposed,
one gets larger values of $\sin 2 \beta$ \cite{reina}.
\begin{figure}   
    \centering
    \epsfxsize=1.08\textwidth
    \leavevmode\epsffile{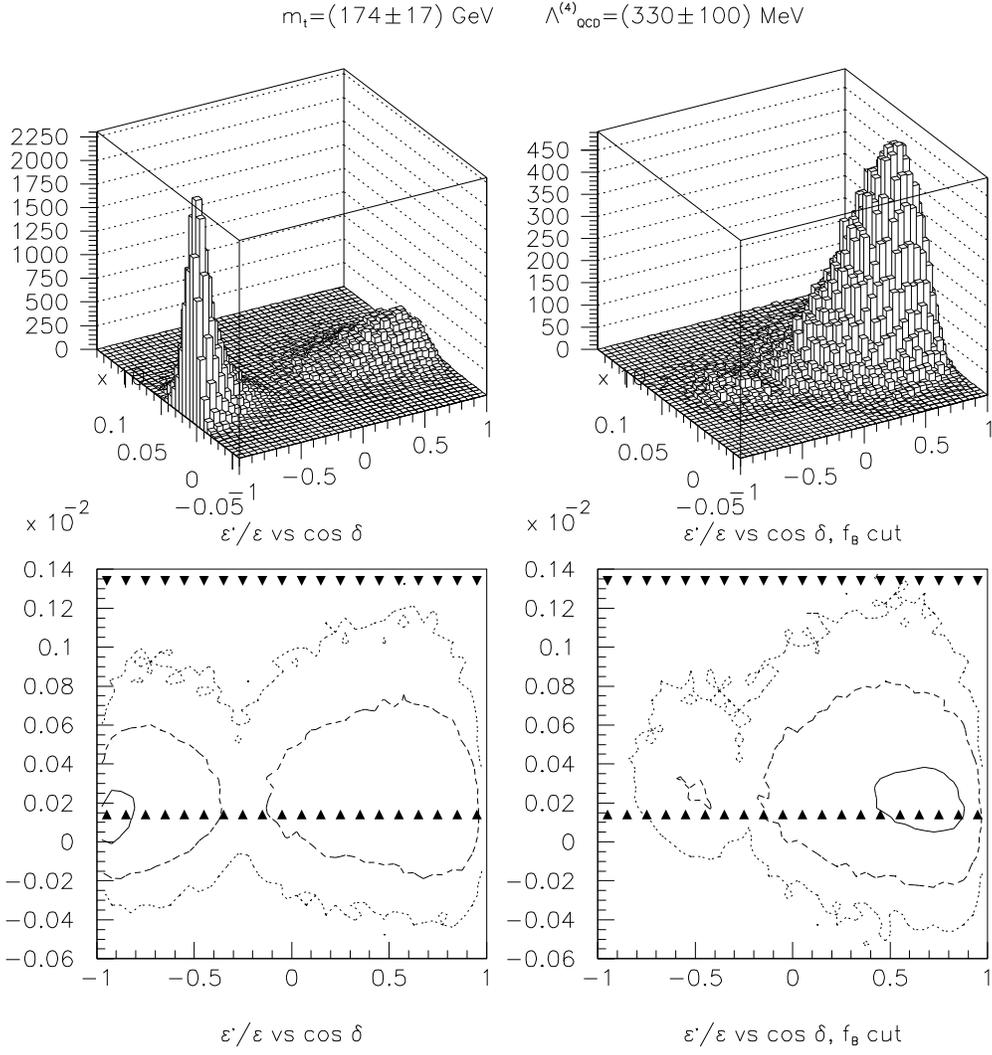}
       \caption[]{\it{Distributions of the events in the
plane $\epse$--$\cos \delta$ without
and with the $f_B$-cut. The corresponding
contour plots are displayed below the lego-plots.}}
\protect\label{fig:tuttoepe}
\end{figure}
{}From the dashed distribution, we obtain
\be \sin 2 \beta=0.65 \pm 0.12\, \, . \protect\label{eq:s2b} \ee
\par d) In fig. \ref{fig:tuttoepe}, several pieces of information
on $\epse$ are provided. Lego-plots of the distribution of the generated
events in the $\epse$--$\cos \delta$ plane are shown, without
and with the $f_B$-cut. At the same time, the corresponding
contour plots are displayed. One notices a very mild dependence
of $\epse$ on $\cos \delta$. As a consequence, one obtains approximately
the same prediction in the two cases (see also fig. \ref{fig:cd}). In
the $HV$ scheme \cite{hv} the results are
\be \epse = (2.3 \pm 2.1 )\times  10^{-4} \,\,\, \hbox{No-cut}, \ee
and
\be \epse = (2.8 \pm 2.4 )\times  10^{-4} \,\,\, f_B\hbox{-cut},
\protect\label{eq:runo}
\ee whereas in the $NDR$ scheme they become
\be \epse = (2.8 \pm 2.2 )\times  10^{-4} \,\,\, \hbox{No-cut}, \ee
and
\be \epse = (3.4 \pm 2.5  )\times  10^{-4} \,\,\, f_B\hbox{-cut}.
\protect\label{eq:rdue}
\ee
Our best estimate, reported in the abstract, has been obtained
by averaging the results given in eqs.(\ref{eq:runo}) and (\ref{eq:rdue})
and by adding a systematic error, due to the difference of the
central values in the two schemes, to the final result
\be  \epse = (3.1 \pm 2.5 \pm 0.3   )\times  10^{-4} \,\,\, f_B\hbox{-cut}. \ee
\par e) In view of the rapid evolution of the experimental determination
of the value of the top mass, expected in the near future,
we also give $\epse$ as a function of $m_t$, in fig. \ref{fig:mtd}.
\par By comparing the present results to the  analysis of ref.
\cite{ciuc1}, one may notice the following:
\begin{enumerate}
\item The latest values of $\vert V_{ub}/V_{cb}\vert$
and $\vert V_{cb} \vert$ are sensibly lower than the previous ones.
Consequently,
the separation of the positive--negative  $\cos \delta$ solutions
takes place at larger values of $m_t$.
\item $\epse \sim \vert V_{ub}/V_{cb}\vert \times \vert V_{cb} \vert^2$.
The actual results for $\epse$, however, are not very different from
 those  given in ref.
\cite{ciuc1}, because we are now using a smaller value
of the  strange quark mass taken
from lattice $QCD$ \cite{ms}, and not from $QCD$ sum rules
\cite{becchi}--\cite{domi2},  as we did before. One has new reasons to doubt
of those $QCD$ sum rule determinations, which have been found
to be affected by large higher
order \cite{larin} or instanton effects \cite{nason}. It is reassuring
that
the recent $QCD$ sum rule calculation of ref. \cite{jamun}, apparently free of
uncontrolled non-perturbative effects, agrees with the lattice calculation
of ref. \cite{ms}.
\item In spite of several differences,
 the bulk of our results overlaps with those of  ref. \cite{burasepe}.
It is encouraging that  theoretical predictions,
obtained by using different approaches  to evaluate
the operator matrix elements, are in good agreement.
\item On the basis of the latest analyses,
 it seems very difficult
that $\epse$ is larger than $10 \times 10^{-4}$. Theoretically,
this may happen  by taking the matrix elements of the
dominant  operators, $Q_6$ and $Q_8$ (see below)
 very different from usually assumed.
One possibility,  discussed in ref. \cite{burasepe}, is to take the
corresponding  $B$-parameters
to be $B_6 \sim 2$ and $B_8 \sim 1$, instead of the  values
$B_6 \sim B_8 \sim 1 $, adopted here and in refs. \cite{reina,ciuc1,burasepe}.
To our knowledge, no coherent
theoretical approach has found $B_6$ so large, and we  see
no reason to use $B_6=2$.
\end{enumerate}
\section{$\ep$, $B^0-\bar B^0$ transitions and $\epp/\ep$}
\protect\label{sec:hs}
\par
In this section, we list the expressions of the
effective Hamiltonians,  responsible for $\DStwo$, $\Delta B=2$ and
$\DSone$ transitions,
given in terms of the relevant operators and their Wilson coefficients.
{}From these  Hamiltonians, one derives the expressions for
$\epsilon$, $x_d$ and
$\epsilon^{\prime}/\epsilon$, written as combinations
of the  Wilson coefficients
times the $B$-parameters of the different operators, i.e. their matrix
elements.
The formulae of this section have been used   in our study.
\par
1)  The effective Hamiltonian governing the $\DStwo$ amplitude is given by
\bea
{\cal H}_{eff}^{|\Delta S|=2}
= \frac{G_{ F}^2}{16{\pi}^2}M_{ W}^2
({\bar d}{\gamma}^{\mu}_{ L}s)^2\left\{{\lambda}_c^2F(x_c)+{\lambda}_t
^2F(x_t)+2{\lambda}_c{\lambda}_tF(x_c,x_t)\right\}, \nonumber \\
& &
\protect\label{effective hamiltonian1}
\eea
where $G_{ F}$ is the Fermi coupling constant and
$\gamma^{\mu}_{ L}=\gamma^{\mu}(1-\gamma_5)\,$;
${\lambda}_q$'s are related to the CKM matrix elements by
${\lambda}_q=V_{qi}V^{\star}_{qf}$,
where `$i$' and `$f$' are the labels of the initial and final states
respectively (in the present case
$i=s$ and $f=d$);
$x_q={m_q^2}/{M_{ W}^2}~$ and
the functions $F(x_i)$ and $F(x_i,x_j)$ are the so-called
{\it Inami-Lim} functions  \cite{inami}, obtained from the calculation of the
basic box-diagram and including $QCD$ corrections  \cite{buras0};
 $F(x_t)$ is known at the
next-to-leading order, which has been included in our calculation.
{}From equation (\ref{effective hamiltonian1}) we can derive the CP violation
parameter $\epsilon$
\bea
|\epsilon|_{\xi=0}=C_{ \epsilon}B_{ K}A^2\lambda^6\sigma\sin\delta
\left\{F(x_c,x_t)+F(x_t)[A^2\lambda^4(1-\sigma\cos\delta)]
-F(x_c)\right\}, \nn \\
\protect\label{epsilon_csizero_1}
\eea
where
\be
C_{ \epsilon}=\frac
{G_{ F}^2f_{ K}^2 M_{ K}M_{ W}^2}{6\sqrt 2{\pi}^2\Delta M_K}.
\ee
Here, $\Delta M_K$ is the mass difference between the two neutral kaon mass
eigenstates.
In eq. (\ref{epsilon_csizero_1}), $\rho=\sigma
\cos \delta$, $\eta=\sigma \sin\delta$ and $\lambda$, $A$, $\rho$ and $\eta$
are the
parameters of the CKM matrix in the Wolfenstein parametrization \cite{wolf};
$B_{ K}$ is the renormalization group-invariant $B$-factor, to be discussed
in sec. \ref{sec:bpar};
$B_K$ takes into account all
the possible deviations from the vacuum insertion approximation in the
evaluation of the matrix element
$\langle\Kbzero|(\bar d\Gmulup s)^{ 2}|\Kzero\rangle$
 ($B_{ K}=1$ corresponding to the vacuum insertion
approximation).
\par 2) The $\Delta B=2$ effective Hamiltonian is given by
\bea
{\cal H}_{eff}^{|\Delta B|=2}
= \frac{G_{ F}^2}{16{\pi}^2}M_{ W}^2 {\lambda}_t^2
({\bar d}{\gamma}^{\mu}_{ L}b)^2
F(x_t), \nonumber \\
& &
\protect\label{effective hamiltonian2}
\eea
from which one finds
\bea x_d &=&\frac{\Delta M}{\Gamma}=
 C_B \frac{\tau_B f_B^2}{M_B} B_B
A^2 \lambda^6 \Bigl( 1 +\sigma^2 -2 \sigma \cos\delta \Bigr)
F(x_t), \nn \\
C_B &=& \frac{G_F^2 M_W^2 M_B^2}{6 \pi^2}, \protect\label{eq:xd} \eea
where $B_B$ is the renormalization-invariant
$B$-parameter, relevant for $B$--$\bar B$ mixing.
 \par 3) The $\Delta S=1$ effective Hamiltonian above the charm threshold
is given by
\bea
{\cal H}_{eff}^{\DSone}&=&\lambda_u \frac {G_F} {\sqrt{2}}
\Bigl[ (1 - \tau ) \Bigl( C_1(\mu)\left( Q_1(\mu) - Q_1^c(\mu) \right) +
C_2(\mu)\left( Q_2(\mu) - Q_2^c(\mu) \right)  \Bigr)\nn\\
&+&\tau \vec Q(\mu)^T \cdot \vec C(\mu) \Bigr]
\protect\label{eh}
\eea
where $\vec Q(\mu)=(Q_1(\mu),Q_2(\mu),\dots)$,
$\vec C(\mu)=(C_1(\mu),C_2(\mu),\dots)$,
$\lambda_u = V_{ud} V^*_{us}$ and similarly
$\lambda_c$ and $\lambda_t$; $\tau=-\lambda_t/\lambda_u$
 and $\mu$ is the renormalization scale of the operators $Q_i$.
A convenient  basis of operators \cite{russi}--\cite{lus},
 when $QCD$ and $QED$ corrections are taken
into account, is
\bea
Q_{ 1}&=&({\bar s}_{\alpha}d_{\alpha})_{ (V-A)}
    ({\bar u}_{\beta}u_{\beta})_{ (V-A)}
   \nn\\
Q_{ 2}&=&({\bar s}_{\alpha}d_{\beta})_{ (V-A)}
    ({\bar u}_{\beta}u_{\alpha})_{ (V-A)}
\nn \\
Q_{ 3,5} &=& ({\bar s}_{\alpha}d_{\alpha})_{ (V-A)}
    \sum_{q}({\bar q}_{\beta}q_{\beta})_{ (V\mp A)}
\nn \\
Q_{ 4,6} &=& ({\bar s}_{\alpha}d_{\beta})_{ (V-A)}
    \sum_{q}({\bar q}_{\beta}q_{\alpha})_{ (V\mp A)}
\nn \\
Q_{ 7,9} &=& \frac{3}{2}({\bar s}_{\alpha}d_{\alpha})_
    { (V-A)}\sum_{q}e_{ q}({\bar q}_{\beta}q_{\beta})_
    { (V\pm A)}
\nn \\
Q_{ 8,10} &=& \frac{3}{2}({\bar s}_{\alpha}d_{\beta})_
    { (V-A)}\sum_{q}e_{ q}({\bar q}_{\beta}q_{\alpha})_
    { (V\pm A)} \nn \\
Q^c_{ 1}&=&({\bar s}_{\alpha}d_{\alpha})_{ (V-A)}
    ({\bar c}_{\beta}c_{\beta})_{ (V-A)}
\nn \\
Q^c_{ 2}&=&({\bar s}_{\alpha}d_{\beta})_{ (V-A)}
    ({\bar c}_{\beta}c_{\alpha})_{ (V-A)}
\protect\label{epsilonprime_basis}
\eea
where the subscript $(V \pm A)$ indicates the chiral structure and
$\alpha$ and $\beta$ are colour indices.
The sum over the quarks $q$       runs over the active flavours at the
scale $\mu$.
\par From ${\cal H}_{eff}^{|\Delta S=1|}$
we can derive the expression for $\epsilon^{\prime}$
\be  \epsilon^{\prime}=\frac{e^{ i\pi/4}}{\sqrt{2}}\frac{\omega}
{\mbox{Re}\,A_{ 0}}\left[\omega^{ -1}
(\mbox{Im}\,A_{ 2})^{\prime}-(1-\Omega_{ IB})\,\mbox{Im}\, A_{ 0}
\right] ,
\protect\label{epsilonprime}
\ee
where
$(\mbox{Im}\,A_{ 2})^{\prime}$ and $\mbox{Im}\,A_{ 0}$ are given by
\bea
\mbox{Im}\,A_{ 0} &=&-\GF \mbox{Im}\,\Bigl({ V}_{ ts}^{ *}{ V}_{ td}\Bigr)
\left\{-\left(C_{ 6}B_{ 6}+\frac{1}{3}C_{ 5}B_{ 5}\right)Z
+\left(C_{ 4}B_{ 4}+\frac{1}{3}C_{ 3}B_{ 3}\right)X+\right.
\nn\\
& &\!C_{ 7}B_{ 7}^{ 1/2}\left(\frac{2Y}{3}+\frac{Z}{6}+
\frac{X}{2}\right)+C_{ 8}B_{ 8}^{ 1/2}\left(2Y+\frac{Z}{2}+
\frac{X}{6}\right)-\nn\\
& &\!\left.C_{ 9}B_{ 9}^{ 1/2}\frac{X}{3}+\left(\frac{C_{ 1}
B_{ 1}^{ c}}{3}+C_{ 2}B_{ 2}^{ c}\right)X\right\},
\protect\label{ima0}
\eea
and
\bea
(\mbox{Im}\, A_{ 2})^{\prime}\!&=&\!-G_{ F}\mbox{Im}\,
\Bigl({ V}_{ ts}^{ *}{ V}_{ td}\Bigr)
\left\{C_{ 7}B_{ 7}^{ 3/2}\left(\frac{Y}{3}-\frac{X}{2}\right)+
\right. \nn \\
& &\!\left.C_{ 8}B_{ 8}^{ 3/2}\left(Y-\frac{X}{6}\right)+
C_{ 9}B_{ 9}^{ 3/2}\frac{2X}{3}\right\}.
\protect\label{ima2}
\eea
$ \omega=\mbox{Re}\,A_{ 2}/ \mbox{Re}\,A_{ 0}=0.045 $ and
we have introduced $( \mbox{Im}\,A_{ 2})^{\prime}$ defined as
\be
\mbox{Im}\,A_{ 2}=(\mbox{Im}\,A_{ 2})^{\prime}+\Omega_{ IB}
(\omega\,\mbox{Im}\,A_{ 0}).
\ee
$\Omega_{ IB}= 0.25 \pm 0.10$
represents the isospin breaking contribution, see for example ref.
\cite{bur5}.
With the subtle cancellations, occurring for a large top mass, $m_t \sim 175$
 GeV, a precise knowledge of the value of $\Omega_{ IB}$ becomes important.
For example, by taking $\Omega_{ IB}=0$, one gets $\epse \sim 5 \times
10^{-4}$,
instead of $2.8 \times 10^{-4}$.  In view of this, further efforts should
be devoted to improve  the accuracy of the theoretical  predictions
for isospin-breaking effects.
\par The numerical evaluation of $\epsilon^{\prime}/\epsilon$ requires the
knowledge  of the Wilson coefficients of the operators and of the
corresponding matrix elements.
The Wilson coefficients have been
evaluated at the next-to-leading order, using the anomalous
dimension matrices given in refs. \cite{alta}--\cite{noi},
and the initial conditions computed in
refs. \cite{flynn1,buras1} (and given for $HV$ and $NDR$ in
refs. \cite{bur2}--\cite{noi}).
The matrix elements of the operators have been written in terms of the
three quantities (see eqs. (\ref{ima0}) and
(\ref{ima2}))
\bea
X\!&=&\!f_{\pi}\left(M_{ K}^{ 2}-M_{\pi}^{ 2}\right), \\
Y\!&=&\!f_{\pi}\left(\frac{M_{ K}^{ 2}}{m_s(\mu)+\md(\mu)}\right)
^2\sim 12\,X\left(\frac{0.15 \, \mbox{GeV}}{m_s(\mu)}\right)^2, \\
Z\!&=&\!4\left(\frac{f_{ K}}{f_{\pi}}-1\right)Y,
\eea
and a set $\{B_{ i}\}$ of $B$-parameters (in our normalization $f_{\pi}
=132$ MeV).
The numerical value of the $B$-parameters have been taken from lattice
calculations \cite{shar3}--\cite{shar4}
 and multiplied by suitable renormalization factors
to take into account the difference between $HV$ ($NDR$) and the lattice
regularization scheme. For those $B$-factors that  have not been computed
yet on the lattice we have used an educated guess, which will be discussed
in detail in the following.
We observe that, in eqs. (\ref{ima0}) and (\ref{ima2}), only nine
coefficients ($B$-parameters) appear since we have used the relation
$ Q_{ 10}=-Q_{ 3} + Q_{ 4} + Q_{ 9}$, which is valid below the bottom
threshold.
\section{Wilson expansion and renormalization}
\protect\label{sec:coe}
This section is devoted to
the Wilson coefficients of the operators
appearing in the effective $\Delta S=1$ Hamiltonian, which we have
computed at the next-to-leading order, including $QCD$ and
$QED$ corrections.
We do not discuss the determination of the coefficients of the
operators relevant for $\epsilon$,
since this point was explained in great detail in
ref. \cite{reina}.
\par The effective Hamiltonian is  defined by the Wilson
Operator Product Expansion (OPE) of the $T$-product of two weak currents:
\bea \langle F \vert {\cal H}_{eff}^{|\Delta S=1|}\vert I
\rangle  &=& g_W^2/8 \int d^4 x
D^{\mu\nu}_W(x^2, M_W^2) \langle F
 \vert T \left( J_{\mu}(x),J^{\dagger}_{\nu} (0)
\right) \vert I \rangle \nn \\ &\rightarrow& \sum_{i}
C_{ i}(M_W) \langle F \vert Q_{ i}(M_W)  \vert I \rangle. \eea
The Wilson coefficients $\vec C(M_W)=(C_1(M_W), C_2(M_W) ...)$
are found by matching, at $O(\alphae)$
and $O(\alphas)$, the current--current and
penguin diagrams computed with the $W$ and top propagators to
those computed with the local four-fermion operators in the effective theory.
The $\vec C(\mu)$ are expressed
in terms of  $\vec C(M_W)$ through the
 renormalization evolution matrix $\W[\mu,M_W]$\footnote{ We have
properly taken into account the beauty threshold in the evolution matrix
\cite{bur2}--\cite{noi}.}
\be
\vec C(\mu) = \W[\mu,M_W] \vec C(M_W), \protect\label{evo} \ee where
\bea
\W[\mu,M_W]   = \hat M[\mu] \U[\mu, M_W] \hat M^{\prime}[M_W],
 \protect\label{monster} \eea
with \be  \hat M[\mu] =
\left(\hat 1 +\frac{\alphae }{4\pi}\Ke\right)
  \left(\hat 1 +\frac{\alphas (\mu)}{4\pi}\J\right)
 \left(\hat 1+\frac{\alphae}{\alphas (\mu)}\PP\right)
\protect\label{mo1} \ee
and
\be \hat M^{\prime}[M_W]=\left(\hat 1-\frac{\alphae}{\alphas (M_W)}\PP\right)
            \left(\hat 1 -\frac{\alphas (M_W)}{4\pi}\J\right)
\left(\hat 1 -\frac{\alphae }{4\pi}\Ke\right).
\protect\label{mo2} \ee

At the next-to-leading accuracy, $\W[\mu,M_W]$ is regularization scheme
dependent.
\par We now give the basic information necessary to compute the
matrices defined in eqs. (\ref{monster}), (\ref{mo1}) and
(\ref{mo2}).
\par The matrix $\U[\mu,M_W]$ in eq. (\ref{monster}) is given by
\be
\U[\mu,M_W]= \left[\frac{\alphas (M_W)}{\alphas (\mu)}\right]^{
            \gammazeros / 2\beta_{ 0}}\, .
\protect\label{u0} \ee
The matrices $\PP$, $\J$ and $\Ke$ are solutions of the equations
\be
\PP+\left[\PP,\frac{\gammazeros}{2\beta_{ 0}}\right] = \frac{\gammazeroe}
   {2\beta_{ 0}} \protect\label{pp} \ee
\be
\J-\left[\J,\frac{\gammazeros}{2\beta_{ 0}}\right] =
         \frac{\beta_{ 1}}{2\beta^2_{ 0}}\gammazeros-
         \frac{\gammaones}{2\beta_{ 0}} \protect\label{jj}
\ee
\be
\left[\Ke,\gammazeros \right]
=\gammaonee+\gammazeroe \J+\gammaones \PP+\left[\gammazeros,\J\PP\right]
    -2\beta_1 \PP -\frac{\beta_{ 1}}{\beta_{ 0}}\PP\gammazeros \, .
\protect\label{ke} \ee
The anomalous dimension matrix, which includes gluon and photon corrections,
can be written as
\be
\hat \gamma= \frac {\alphas }{ 4 \pi } \hat \gamma_s^{(0)} +
 \frac {\alphae }{4 \pi} \hat \gamma_e^{(0)}
+ \Bigl(\frac {\alphas }{4 \pi}\Bigr)^2 \hat \gamma_s^{(1)} +
 \frac{ \alphas }{4 \pi} \frac{ \alphae}{4 \pi}  \hat \gamma_e^{(1)}
\, , \ee
where each of the $\hat \gamma^{(0,1)}_{s,e}$ is a $10 \times 10$ matrix.
In
eqs. (\ref{u0})--(\ref{ke}), $\beta_{ 0}$ and $\beta_{ 1}$ are the
first two coefficients of the $\beta$-function of $\alphas$.
At the leading order, the $QCD$ anomalous dimension matrix,
including $QCD$ penguins, has been
computed in refs. \cite{russi,gilman}. The electroweak anomalous dimension
matrix at the same order can be found in refs. \cite{bij},
\cite{lus} and  \cite{buras1}.
Two groups have computed
the anomalous dimension matrix at the next-to-leading order,
  by calculating all the current--current and penguin
operators at two loops up to order $\alphas^2 t$  and $\alphae \alphas t$
\cite{bur2}--\cite{noi}. All the details of the computation can
 be found in these references. Here
 we only give the numerical results for the Wilson coefficients
in three different
regularization/renormalization schemes, which are commonly used in the
literature, namely $HV$, $NDR$ and $RI$:
\begin{enumerate}
\item
$HV$ denotes the $\overline{MS}$ renormalization prescription in
dimensional regularization, with $\gamma_5$ defined according to the
't Hooft and Veltman prescription \cite{hv};
\item $NDR$ is the $\overline{MS}$ scheme
 in Na\"\i ve Dimensional Regularization,
with anticommuting $\gamma_5$;
\item The $RI$ (Regularization Independent) scheme requires a few
words of explanation. In its simplest realization,
one fixes the renormalization conditions of a
certain operator, by imposing that suitable Green functions, computed
between external off-shell quark and gluon states, in a fixed gauge,
coincide with their tree level value. For example, if we consider
the generic two-quark operator $O_\Gamma= \bar \psi \Gamma \psi$,
we may impose the condition
\be Z_\Gamma \langle p \vert O_\Gamma \vert p \rangle\vert_{p^2=-\mu^2}=
\langle p \vert O_\Gamma \vert p \rangle_0, \ee
where $\Gamma$ is one of the Dirac matrices and $
\langle p \vert O_\Gamma \vert p \rangle_0$ is the tree level matrix element.
The extension to more complicated cases, including four-fermion operators
and operator mixing, is straightforward and will be discussed
in detail in  sec. \ref{sec:ri}.
The $RI$ procedure defines the same renormalized operators, i.e. the same
coefficients, in all  regularization schemes ($HV$, $NDR$, Pauli--Villars,
lattice), provided they are expressed in terms of the same renormalized
strong coupling constant. However, the coefficient functions now depend on
the external states and on the gauge used to impose the renormalization
conditions\footnote{The ``physical", hadronic
matrix elements of  the effective Hamiltonian are independent of the
external
states and of the gauge used to fix the renormalization conditions up
to the order at which one is working, in our case up to the NLO.}.
Thus the external states and the gauge must be specified.
 To make contact with the non-perturbative
method proposed in ref. \cite{nonper}, we will use the Landau gauge.
The external states will be different for different
operators, and will be
given in sec. \ref{sec:ri}.
\par  The $RI$ scheme is particularly convenient for
matching the coefficients of the operators to
the corresponding matrix elements computed in lattice simulations,
as extensively discussed in ref. \cite{nonper}.
The details of the implementation of this renormalization scheme
will be given in sec. \ref{sec:ri}, where we also present the relations
between the renormalized operators and coefficients computed with the
RI scheme and with the usual $\overline{MS}$ prescriptions.
\end{enumerate}
\section{The effective Hamiltonian written  in terms of lattice operators}
\protect\label{sec:latcon}
Hadronic matrix elements from lattice $QCD$ have been widely used to
predict several quantities of phenomenological
interest in weak decays, deep inelastic scattering
and heavy flavour physics.
In this section we discuss the renormalization properties
of lattice operators and clarify some issues related
to the construction of physical amplitudes  starting from
the matrix elements computed on the lattice.
For definiteness and simplicity, we will refer to
two-fermion operators and to the four-fermion $\Delta I=3/2$
operator $Q^+$, which renormalize multiplicatively in the
continuum\footnote{$Q^+$ is related, via a chiral transformation, to the
 $\Delta S=2$ operator of the effective Hamiltonian,
 introduced in eq. (\ref{effective hamiltonian1}).
We have chosen to discuss $Q^+$ to avoid the complications
present in the calculation of the effective $\Delta S=2$
Hamiltonian, due to the
matching conditions in the presence of a heavy top quark.}.
 The relations
derived below are, however, valid  in the general case.
These relations have been used to compute  the strange
and  charm  quark masses in ref. \cite{ms}.
\par
Let us consider first the renormalization of the generic two-quark operator
$\bar \psi \Gamma \psi$, where  $\Gamma$ is one of the Dirac
matrices. At the next-to-leading order (NLO),
the generic, forward, two-point Green function,
  computed between quark states of virtuality $p^2=\mu^2$, has the
form\footnote{We work on
an Euclidean lattice.}
\bea \Gamma(\mu a)& =&
 \Gamma_0 \left\{ 1 + \frac{g^2_L(a)}{16 \pi^2}
\left(\frac{ \gamma^{(0)}}{2} \ln\left( \frac{\mu a }{\pi}\right)^2 +C^L
\right)
\right. \nn \\
&+&\Bigl(\frac{g^2_L(a)}{16 \pi^2}\Bigr)^2
\left[ \frac{1}{8} \gamma^{(0)}
( -2 \beta_0 + \gamma^{(0)})
\ln^2 \left(\frac{\mu a }{\pi}\right)^2 \right. \\
 &+& \left.\left.
\frac{1}{2} \left( \bar \gamma^{(1)}+
 ( -2 \beta_0 + \gamma^{(0)}) C^L
+\beta^0_\lambda \left(\lambda \frac{\partial C^L}{\partial \lambda}
\right) \right) \ln \left(\frac{\mu a }{\pi}\right)^2
\right] \right\}  + \dots  \nn \protect\label{eq:rgel} \eea
where the dots represent terms beyond the next-to-leading order and
terms of $O(a)$, which we will assume negligible in the following;
$\Gamma_0$ is the zeroth-order Green function, and  $g^2_L(a)=6/\beta$ is the
bare lattice coupling constant; $\gamma^{(0)}$ and
$\bar \gamma^{(1)}$ are the leading (regularization-independent) and
next-to-leading (regularization-dependent)
order anomalous dimensions, respectively;
$\lambda=\lambda(a)$ is the lattice gauge parameter
of the gluon propagator $\Bigl( \Pi_{\mu \nu}(q^2)= -\delta_{\mu \nu}
+(1-\lambda) q_\mu q_\nu/q^2 \Bigr)$. It obeys the renormalization
group equation
\be \frac{1}{\lambda(a)}  a \frac{\partial \lambda(a)}
{\partial a}=-\beta_\lambda(g^2_L(a))=
-\beta^0_\lambda \frac{g^2_L(a)}{16\pi^2}+\cdots, \,\,\,\,
\,\,\,\,\, \beta^0_\lambda=\frac{5 N-2 n_f}{6\pi}. \protect\label{eq:bl} \ee
We have introduced a scale-dependent gauge parameter in order to define a
gauge-independent anomalous dimension and simplify the renormalization
group equation for $\Gamma(\mu a)$, see eq. (\ref{eq:rga})  below.
\par The lattice coupling constant obeys the equation
\be a \frac { dg^2_L(a) }
{d a} =- \beta (g^2_L(a) )= 2 \beta_0 \frac{g^4_L(a)}{16\pi^2}
 +2 \beta_1 \frac{g^6_L(a)}{(16\pi^2)^2}, \protect\label{eq:rcc}\ee
where $\beta_{0,1}$ are given by
\be \beta_0 =\frac {(11N-2 n_f)} {3},
\,\,\,\,\,\,\,\,\,\,\, \beta_1 = \frac {34}{3} N^2 - \frac{10}{3} N n_f -\frac
{(N^2-1)}{N} n_f \ee
and $n_f$ is the number of flavours.
Equation (34) guarantees that all the  matrix elements
can be made finite
(as $a \rightarrow 0$) by  multiplying  the bare operator
by a suitable renormalization
constant, obtained by   fixing  the renormalization conditions
for $O_\Gamma$.

\par From the above equations, we find
\be \Bigl( a \frac{\partial}{\partial a} -
\beta(g^2_L(a))\frac {\partial}{\partial g^2_L(a)}-\beta_\lambda(g^2_L(a))
\lambda\frac{\partial}{\partial \lambda}
 -\bar \gamma(g^2_L(a)) \Bigr) \Gamma(\mu a) =0 ,
\protect\label{eq:rga} \ee
with
\be \bar \gamma(g^2_L(a)) =  \frac{g^2_L(a)}{16\pi^2} \gamma^{(0)}
 + \frac{g^4_L(a)}{(16\pi^2)^2}\bar \gamma^{(1)}. \protect\label{gcc}\ee
\par In view of the comparison with some continuum regularization,
it is convenient to expand the bare Green function $\Gamma(\mu a)$
of eq. (34) in terms of the continuum minimal subtraction
($\overline{MS}$)  coupling constant, evaluated at the scale $\pi/a$.
The continuum $\overline{MS}$ coupling  $\as(\pi/a)$ is related to
 the lattice bare  coupling
$\as^L(a)=g^2_L(a)/4 \pi$ by the equation
\be
\frac{1}{\as^L(a)}=\frac{1}{\as(\pi/a)} \Bigl( 1 + \frac{\as(\pi/a)}{4\pi}
\Delta+\dots \Bigr), \protect\label{eq:ras} \ee where $\Delta$ is a numerical
constant.
With this substitution, eq. (\ref{eq:rga}) becomes
\be \Bigl( a
\frac{\partial}{\partial a} - \beta(\as)\frac {\partial}{\partial
\as}-\beta_\lambda(\as) \lambda\frac{\partial}{\partial \lambda}
 -\gamma_L(\as) \Bigr) \Gamma(\mu a) =0
\protect\label{eq:rga1} \ee
and
\be \beta(\as) = - 2 \beta_0 \frac{\as^2}{4\pi}
 -2 \beta_1  \frac{\as^3}{(4\pi)^2}, \,\,\,\,\,\,\,\,\,\,
 \gamma_L(\as) =  \frac{\as}{4\pi}\gamma^{(0)}
 +  \frac{\as^2}{(4\pi)^2}\gamma^{(1)}_L, \protect\label{eq:gccb}\ee
\be \beta_\lambda(\as)=\beta^0_\lambda \frac{\as}{4 \pi}+\cdots  \nn
\ee
By changing the expansion parameter, one also has to change the
two-loop anomalous dimension \cite{alta}--\cite{noi}, see also
eq. (34)
\be \gamma^{(1)}_L=\bar \gamma^{(1)} -  \Delta \gamma^{(0)}.
\protect\label{eq:rg1} \ee
Finally, the running coupling constant $\as$ is given by
\bea \frac
{\as(\mu^2)}{4 \pi} = \frac {1} {\beta_0 \ln(\mu^2/\Lambda_{QCD}^2)} \Bigl(
1 - \frac{\beta_1 \ln[\ln(\mu^2/\Lambda_{QCD}^2)]}{\beta_0^2
\ln(\mu^2/\Lambda_{QCD}^2)}\Bigr) + \cdots \protect\label{eq:srcc} \eea
The above equation defines
the continuum $\overline{MS}$ scale parameter $\Lambda_{QCD}$ at the NLO.

\par In the case of the $\Delta I=3/2$ lattice operator, the coefficient
$C^L$ and the anomalous dimension $\gamma^{(1)}_L$
become a vector  due to the mixing
of the bare operator with operators of different chirality, induced
by the Wilson term present in the lattice   action \cite{mart1,ber1,boc}.
 Operators of different
chirality correspond, in the language of refs. \cite{alta}--\cite{noi},
to the so-called ``effervescent" operators.
Denoting by  $\Gamma_0^1$ the zeroth-order Green
function of the original operator and by  $\Gamma_0^i$ with
$i=2, \dots 4$ the zeroth-order Green functions of the ``effervescent"
operators \cite{mart1,ber1},
eq. (34) becomes
\bea & \, & \Gamma(\mu a) = \sum_{i=1,4}
 \Gamma_0^i \left\{ \delta_{i,1} + \frac{g^2_L(a)}{16 \pi^2}
\left(\delta_{i,1}\frac{ \gamma^{(0)}}{2} \ln \left(
\frac{\mu a }{\pi}\right)^2 +C^L_i \right) \right. \nn \\
&+& \Bigl(\frac{g^2_L(a)}{16 \pi^2}\Bigr)^2
 \left[\delta_{i,1} \frac{1}{8} \gamma^{(0)}
( -2 \beta_0 + \gamma^{(0)})
\ln^2 \left(\frac{\mu a }{\pi}\right)^2 +
\frac{1}{2}\left(  (\gamma^{(1)}_L)_i \right. \right.\nn \\  &+&
\left. \left.
\left. ( -2 \beta_0 + \gamma^{(0)}) C^L_i
+\beta^{(0)}_\lambda \left(\lambda \frac{\partial C^L_i}{\partial \lambda}
\right) \right)\ln\left( \frac{\mu a }{\pi}\right)^2 \right]\right\}
  + \cdots \protect\label{eq:rgel1} \eea
 with $\lambda \partial C^L_i/\partial \lambda \sim \delta_{i,1}$.
\par In the continuum, if we do not remove the $W$ propagator and expand
the four-fermion amplitude in powers of $\as(M_W)$, we do not need to introduce
any regularization. In other words, the only  divergences
appearing in the four-fermion amplitude are those due to the renormalization
of $\as$. Since the $W$-propagator acts as an ultraviolet cut-off
for the divergences of the Green function, we will use in the following
the name $M_W$-regularization, as much as we use the names
dimensional or lattice regularization.
 We thus can write
\bea&\,& \Gamma(\frac{\mu}{M_W} ) =
 \Gamma_0 \left\{ 1 + \frac{\as(M_W)}{4 \pi}
\left(\frac{ \gamma^{(0)}}{2} \ln\left( \frac{\mu  }{M_W}\right)^2 +C^C
\right)\right. \nn \\
&+&\Bigl(\frac{\as(M_W)}{4 \pi}\Bigr)^2
\left[ \frac{1}{8} \gamma^{(0)}
( -2 \beta_0 + \gamma^{(0)})
\ln^2\left( \frac{\mu  }{M_W}\right)^2 +
\frac{1}{2}\left(  \gamma^{(1)}\right. \right. \nn \\ &+&
\left. \left. \left. ( -2 \beta_0 + \gamma^{(0)}) C^C
+\beta^{(0)}_\lambda \left(\lambda \frac{\partial C^C}{\partial \lambda}
\right) \right) \ln\left( \frac{\mu  }{M_W}\right)^2 \right] \right\}
  + \dots \, , \protect\label{eq:rgec} \eea
where $\lambda=\lambda(M_W)$.
$\Gamma(\mu/M_W )$ obeys the renormalization group equation
\be \Bigl(
\frac{\partial}{\partial t} + \beta(\as)\frac {\partial}{\partial \as}
+ \beta_\lambda(\as)\lambda\frac {\partial}{\partial \lambda}
+\gamma(\as) \Bigr) \Gamma \left( t\right) =0,  \protect\label{eq:rga2} \ee
with
\be t= \ln \Bigl( \frac{M_W}{\mu} \Bigr), \ee
\be \gamma(\as) =  \frac{\as}{4\pi}\gamma^{(0)}
 +  \frac{\as^2}{(4\pi)^2}\gamma^{(1)}\protect\label{gcc2}\nn \ee
and
\be  \frac{\partial\as(t)}{\partial t}  =\, -\beta(\as(t)) ,
\,\,\,\,\,\,\,\,\,\,\, \frac{1}{\lambda(t)}
 \frac{\partial\lambda(t)}{\partial t}=\, - \beta_\lambda(\as). \ee
By solving the renormalization group equation for $\Gamma(\mu/M_W)$,
one finds
\bea \Gamma\Bigl(\frac{\mu}{M_W} \Bigr)& = &\Gamma_0
\Bigl(\frac{\as(M_W)}{\as(\mu)} \Bigr)^{\gamma^{(0)}/2 \beta_0}
\Bigl[ 1 +\frac{\as(\mu)}{4\pi} \Bigl( C^C-\frac{\gamma^{(1)}\beta_0-
\gamma^{(0)}\beta_1}{2 \beta_0^2}\Bigr) \nn \\ &+&
\frac{\as(M_W)}{4\pi} \Bigl( \frac{\gamma^{(1)}\beta_0-
\gamma^{(0)}\beta_1}{2 \beta_0^2} \Bigr) \Bigr], \protect\label{eq:solc} \eea
where $C^C=C^C\Bigl(\lambda(\mu)\Bigr)$.
A similar solution is found for $\Gamma(\mu a)$, where
we encounter the complication due
to the mixing with the effervescent operators
\bea \Gamma(\mu a )
=\sum_{i=1,4} \Gamma_0^i \Bigl(\frac{\as(\pi/a)}{\as(\mu)}
\Bigr)^{\gamma^{(0)}/2 \beta_0} \Bigl[ \delta_{i,1} +\frac{\as(\mu)}{4\pi}
\Bigl( C^L_i \nn \eea \bea  -\frac{(\gamma^{(1)}_L)_i\beta_0- \delta_{i,1}
\gamma^{(0)}\beta_1}{2 \beta_0^2}\Bigr)  +
\frac{\as(\pi/a)}{4\pi} \Bigl( \frac{(\gamma^{(1)}_L)_i\beta_0-
\delta_{i,1}\gamma^{(0)}\beta_1}{2 \beta_0^2} \Bigr) \Bigr].
\protect\label{eq:sola} \eea
The comparison of eqs. (\ref{eq:solc}) and (\ref{eq:sola}) allows us
to write the effective Hamiltonian in terms of the lattice bare operators
\be
H_{eff}= \sum_{i=1,4} O_i(a) C_i(M_W a), \ee where
\bea C_i(M_W a )& =&
\Bigl(\frac{\as(M_W)}{\as(\pi/a)} \Bigr)^{\gamma^{(0)}/2 \beta_0}
\Bigl[ \delta_{i,1} +\frac{\as(M_W)-\as(\pi/a)}{4\pi} \Bigl(
\frac{(\gamma^{(1)}_L)_i\beta_0- \delta_{i,1}
\gamma^{(0)}\beta_1}{2 \beta_0^2}\Bigr)  \nn \\&+&
\frac{\as(M_W)}{4\pi} \Bigl(\delta_{1,i} C^C-C^L_i
 \Bigr) \Bigr]. \protect\label{eq:cmw} \eea
Another convenient expression for $C_i(M_W a)$ is given by
\bea C_i(M_W a )& =&
\Bigl(\frac{\as(M_W)}{\as(\pi/a)} \Bigr)^{\gamma^{(0)}/2 \beta_0}
\Bigl[ \delta_{i,1}\Bigl(1 +\frac{\as(M_W)-\as(\pi/a)}{4\pi}
 \frac{\gamma^{(1)}\beta_0-
\gamma^{(0)}\beta_1}{2 \beta_0^2}\Bigr)  \nn \\ &+&
\frac{\as(\pi/a)}{4\pi} \Bigl(\delta_{1,i}
 C^C-C^L_i \Bigr) \Bigr]  \nn \\ &\simeq&
\Bigl(\frac{\as(M_W)}{\as(\pi/a)} \Bigr)^{\gamma^{(0)}/2 \beta_0}
\Bigl(1 +\frac{\as(M_W)-\as(\pi/a)}{4\pi}
 \frac{\gamma^{(1)}\beta_0-
\gamma^{(0)}\beta_1}{2 \beta_0^2}\Bigr)  \nn \\ &\times&
\Bigl[ \delta_{i,1}+\frac{\as(\pi/a)}{4\pi} \Bigl(\delta_{1,i}
 C^C-C^L_i \Bigr) \Bigr] .
 \protect\label{eq:cmw1} \eea
Equation (\ref{eq:cmw1}) has been obtained using the universality of the
combination\footnote{This relation is
true only when we use the same coupling constant on the lattice and in
 the continuum.}
\be \frac{\gamma^{(1)}}{2 \beta_0}-C^C=
\frac{\gamma^{(1)}_L}{2 \beta_0}-C^L. \ee
Equation (\ref{eq:cmw}) is interpreted as follows: the continuum operator,
defined in the
$M_W$-regularization
is matched into the lattice operator at the $M_W$ scale through the
factor $\left( 1+\as(M_W)/4\pi \Bigl( C^C-C^L \Bigr)\right) $  and then
evolved,
according to eq. (\ref{eq:rga1}), from the scale $M_W$ down to
$\pi/a$ \cite{bur2}--\cite{noi}.
On the other hand, eq. (\ref{eq:cmw1}) is understood in the standard lattice
language as follows: the lattice operator,
is matched into the continuum operator at the scale $\pi/a$ through the
factor $\left(1+\as(\pi/a)/4\pi \Bigl( C^C-C^L \Bigr)\right) $  and then
evolved
in the continuum,
according to eq. (\ref{eq:rga2}), from the scale $\pi/a$ up to
$M_W$ \cite{mart1,ber1}.
Equations (\ref{eq:cmw}) and (\ref{eq:cmw1}) apply with trivial changes
to two-quark operators that renormalize multiplicatively on the lattice,
or to operators that mix under renormalization also in the
continuum.
A few comments may be useful at this point:
\begin{itemize}
\item The continuum,  two-loop anomalous dimension
in the $M_W$-regularization $\gamma^{(1)}$
can be obtained from the calculation done  in  the
$NDR$, $HV$ or $DRED$ regularizations \cite{alta}--\cite{noi}, which
we denote as $\gamma^{(1)}_{\overline{MS}}$ ($C_{\overline{MS}}$),
 using the relation
\be \frac{\gamma^{(1)}}{2 \beta_0}-C^C=
\frac{\gamma^{(1)}_{\overline{MS}}}{2 \beta_0}-C_{\overline{MS}}. \ee
\item In eq. (\ref{eq:cmw1}), the coefficients $C^L_i$  can be
 eliminated by defining a new lattice operator
\be O^{\prime}(a)=O_1(a)-\frac{\as(\pi/a)}{4\pi} \sum_{i=1,4} C^L_i
O_i(a). \protect\label{eq:prime}\ee This is equivalent
to the regularization-independent definition
of the  renormalized operators discussed in refs. \cite{alta}--\cite{noi}
and in sec. \ref{sec:ri}.
This definition is such that $\Gamma(\mu=\pi/a)=\Gamma_0^1$
\footnote{ This is not in contrast with
the condition $p^2 \ll (\pi/a)^2$ that we have to impose in order
to avoid discretization errors. The renormalization condition is simply
a consequence of eq. (34), which is derived by expanding
the Green function in $\alpha_s(\pi/a)$,
instead of $g^2_L(a)/4 \pi$,
 with $p^2 \ll (\pi/a)^2$.}. We stress again
that this procedure requires the  knowledge of the external states for
which we have computed $C^L$ and it is in general gauge-dependent.
\item We have decided to expand the lattice Green function in $\as(\pi/a)$.
However, we have the freedom to expand in $\as(1/a)$ or any other scale
we like, since the change will be  completely  compensated at the
NLO \footnote{ This is true provided that the scale that we choose is not so
different from $1/a$ as  to generate new large logarithms.}.
For the same reason we can expand the Green functions
in a different coupling constant,
for example the ``boosted"  coupling $\as^V$ defined
in refs. \cite{lm1}.  There it
was argued that the series in $\as^V$ may  minimize $O(\as^2)$ NNLO
corrections. If we expand in $\as^V$, we have accordingly
to reorganize the expression  used for the coefficient function.
It is wise however to check the stability of the physical amplitudes under
a change in the scale used for $\as$ and assume this as a theoretical
uncertainty.
\item From eq. (\ref{eq:cmw1}) it is possible to give the expression
of the operator from which we can immediately derive the renormalization
group-invariant $B_K$ parameter. This is obtained by using the  operator
\be O_{RGI} = C(a) \bar O(a), \protect\label{eq:rgibk} \ee
with
\be \bar O(a)=O_1(a)\Bigl(1+
\frac{\as(\pi/a)}{4\pi} C^C\Bigr)-\frac{\as(\pi/a)}{4\pi} \sum_{i=1,4} C^L_i
O_i(a)
\nn \ee
or, in the $\overline{{\rm MS}}$ scheme,
\be \bar O(a)^{\overline{MS}}=O_1(a)\Bigl(1+
\frac{\as(\pi/a)}{4\pi} C^{\overline{{\rm MS}}}\Bigr)
-\frac{\as(\pi/a)}{4\pi} \sum_{i=1,4} C^L_i O_i(a)
\nn \ee and
 \be C( a ) =
\Bigl(\as(\pi/a) \Bigr)^{-\gamma^{(0)}/2 \beta_0}
\Bigl[ 1 -\frac{\as(\pi/a)}{4\pi} \Bigl(
 \frac{(\gamma^{(1)}\beta_0-
\gamma^{(0)}\beta_1}{2 \beta_0^2}\Bigr) \Bigr]\, ; \nn \ee
$C^{\overline{{\rm MS}}}$ can be found in refs. \cite{alta,bur2}.
\item Contrary to what  is stated in ref. \cite{shapat}, it is not necessary
to first match  the lattice operators
 to the corresponding operators in
some con\-ti\-nuum renormalization scheme at a low
scale $\mu_0$, with the $W$ integrated out, and then evolve to $M_W$. The
reason
is that, as demonstrated above, the theory with the $W$ does not require
any renormalization when expanded in $\as(M_W)$. Thus we can avoid
the introduction of $\mu_0$ and relate directly the bare lattice operators
to the full theory. The standard approach to evaluate the
theoretical uncertainty, coming from the matching of the lattice to
the continuum, is obtained by varying $\mu_0$ and the effective (``boosted")
lattice coupling constant, used in the perturbative expansion. We have
shown that, indeed, the uncertainty depends only on the effective coupling,
since in our formulae $\mu_0$ never appears. One can use the operator
$O^\prime(a)$ of eq. (\ref{eq:prime}),
computed using   the non-perturbative procedure of ref. \cite{nonper}.
In this case, the matching can be entirely done in the continuum, so that the
 large
corrections, present in  higher orders in the lattice coupling, are eliminated.
\end{itemize}

\section{The $RI$ renormalization scheme}
\protect\label{sec:ri}
The details of the renormalization of the operators and of the calculation
of the Wilson coefficients in $HV$ and $NDR$ have  been extensively discussed
in
the literature \cite{bur2}--\cite{noi}.
Here, we give the main formulae, necessary to renormalize
the operators in the $RI$ renormalization scheme. Using these formulae,
the reader can obtain the renormalized operators, starting from any
regularized version of $QCD$. The matrix elements of the $RI$ renormalized
operators can be combined directly with  the coefficients given in
tables 4  and 5.
{\scriptsize
\begin{table}
\begin{center}
\begin{tabular}{|c|c|c|}\hline
 & Feynman RI & Landau RI \\\hline\hline
& \multicolumn{2}{c|}{$\mu=1.5$ GeV}\\\hline
$C_{1}$ & $(-2.72\pm 0.29\pm 0.00\pm 0.06)\cdot 10^{-1}$
& $(-3.33\pm 0.44\pm 0.00\pm 0.04)\cdot 10^{-1}$
\\
$C_{2}$ & $(108.70\pm 0.73\pm 0.00\pm 0.25)\cdot 10^{-2}$
& $(11.59\pm 0.24\pm 0.00\pm 0.06)\cdot 10^{-1}$
\\
$C_{3}$ & $(2.05\pm 0.27\pm 0.00\pm 0.11)\cdot 10^{-2}$
& $(2.35\pm 0.38\pm 0.00\pm 0.07)\cdot 10^{-2}$
\\
$C_{4}$ & $(-5.60\pm 0.68\pm 0.01\pm 0.10)\cdot 10^{-2}$
& $(-5.93\pm 0.81\pm 0.01\pm 0.05)\cdot 10^{-2}$
\\
$C_{5}$ & $(10.00\pm 0.91\pm 0.01\pm 0.50)\cdot 10^{-3}$
& $(11.54\pm 0.40\pm 0.01\pm 0.42)\cdot 10^{-3}$
\\
$C_{6}$ & $(-1.17\pm 0.30\pm 0.01\pm 0.07)\cdot 10^{-1}$
& $(-1.12\pm 0.28\pm 0.01\pm 0.06)\cdot 10^{-1}$
\\
$C_{7}$ & $(0.02\pm 0.07\pm 0.18\pm 0.02)\cdot 10^{-3}$
& $(0.01\pm 0.07\pm 0.19\pm 0.02)\cdot 10^{-3}$
\\
$C_{8}$ & $(1.68\pm 0.48\pm 0.30\pm 0.08)\cdot 10^{-3}$
& $(1.56\pm 0.44\pm 0.28\pm 0.09)\cdot 10^{-3}$
\\
$C_{9}$ & $(-7.43\pm 0.21\pm 0.75\pm 0.14)\cdot 10^{-3}$
& $(-7.54\pm 0.19\pm 0.76\pm 0.13)\cdot 10^{-3}$
\\
& \multicolumn{2}{c|}{$\mu=2.0$ GeV}\\\hline
$C_{1}$ & $(-2.27\pm 0.22\pm 0.00\pm 0.04)\cdot 10^{-1}$
& $(-2.75\pm 0.32\pm 0.00\pm 0.03)\cdot 10^{-1}$
\\
$C_{2}$ & $(106.88\pm 0.60\pm 0.00\pm 0.21)\cdot 10^{-2}$
& $(11.27\pm 0.17\pm 0.00\pm 0.04)\cdot 10^{-1}$
\\
$C_{3}$ & $(1.73\pm 0.20\pm 0.01\pm 0.07)\cdot 10^{-2}$
& $(1.93\pm 0.26\pm 0.01\pm 0.04)\cdot 10^{-2}$
\\
$C_{4}$ & $(-4.84\pm 0.52\pm 0.00\pm 0.06)\cdot 10^{-2}$
& $(-5.07\pm 0.59\pm 0.01\pm 0.03)\cdot 10^{-2}$
\\
$C_{5}$ & $(10.31\pm 0.13\pm 0.02\pm 0.31)\cdot 10^{-3}$
& $(11.43\pm 0.26\pm 0.02\pm 0.25)\cdot 10^{-3}$
\\
$C_{6}$ & $(-0.90\pm 0.18\pm 0.00\pm 0.04)\cdot 10^{-1}$
& $(-0.86\pm 0.16\pm 0.00\pm 0.04)\cdot 10^{-1}$
\\
$C_{7}$ & $(0.00\pm 0.04\pm 0.19\pm 0.02)\cdot 10^{-3}$
& $(-0.01\pm 0.04\pm 0.20\pm 0.02)\cdot 10^{-3}$
\\
$C_{8}$ & $(1.26\pm 0.27\pm 0.23\pm 0.08)\cdot 10^{-3}$
& $(1.18\pm 0.25\pm 0.21\pm 0.09)\cdot 10^{-3}$
\\
$C_{9}$ & $(-7.71\pm 0.15\pm 0.78\pm 0.16)\cdot 10^{-3}$
& $(-7.81\pm 0.14\pm 0.79\pm 0.16)\cdot 10^{-3}$
\\
\hline\hline
& \multicolumn{2}{c|}{$\mu=3.0$ GeV}\\\hline
$C_{1}$ & $(-1.74\pm 0.15\pm 0.00\pm 0.03)\cdot 10^{-1}$
& $(-2.11\pm 0.21\pm 0.00\pm 0.02)\cdot 10^{-1}$
\\
$C_{2}$ & $(104.70\pm 0.38\pm 0.00\pm 0.18)\cdot 10^{-2}$
& $(10.92\pm 0.10\pm 0.00\pm 0.03)\cdot 10^{-1}$
\\
$C_{3}$ & $(1.37\pm 0.13\pm 0.01\pm 0.05)\cdot 10^{-2}$
& $(1.50\pm 0.16\pm 0.01\pm 0.03)\cdot 10^{-2}$
\\
$C_{4}$ & $(-3.97\pm 0.37\pm 0.00\pm 0.04)\cdot 10^{-2}$
& $(-4.12\pm 0.41\pm 0.00\pm 0.02)\cdot 10^{-2}$
\\
$C_{5}$ & $(9.78\pm 0.35\pm 0.03\pm 0.18)\cdot 10^{-3}$
& $(10.53\pm 0.52\pm 0.03\pm 0.13)\cdot 10^{-3}$
\\
$C_{6}$ & $(-6.49\pm 0.96\pm 0.03\pm 0.25)\cdot 10^{-2}$
& $(-6.29\pm 0.91\pm 0.03\pm 0.21)\cdot 10^{-2}$
\\
$C_{7}$ & $(0.00\pm 0.02\pm 0.19\pm 0.02)\cdot 10^{-3}$
& $(0.01\pm 0.02\pm 0.20\pm 0.02)\cdot 10^{-3}$
\\
$C_{8}$ & $(0.90\pm 0.15\pm 0.16\pm 0.08)\cdot 10^{-3}$
& $(0.84\pm 0.14\pm 0.15\pm 0.08)\cdot 10^{-3}$
\\
$C_{9}$ & $(-7.93\pm 0.11\pm 0.81\pm 0.18)\cdot 10^{-3}$
& $(-8.01\pm 0.10\pm 0.82\pm 0.18)\cdot 10^{-3}$
\\
\hline\hline
& \multicolumn{2}{c|}{$\mu=4.0$ GeV}\\\hline
$C_{1}$ & $(-1.41\pm 0.12\pm 0.00\pm 0.02)\cdot 10^{-1}$
& $(-1.73\pm 0.16\pm 0.00\pm 0.01)\cdot 10^{-1}$
\\
$C_{2}$ & $(103.40\pm 0.26\pm 0.00\pm 0.17)\cdot 10^{-2}$
& $(107.32\pm 0.74\pm 0.00\pm 0.29)\cdot 10^{-2}$
\\
$C_{3}$ & $(11.80\pm 0.98\pm 0.14\pm 0.35)\cdot 10^{-3}$
& $(1.27\pm 0.12\pm 0.01\pm 0.02)\cdot 10^{-2}$
\\
$C_{4}$ & $(-3.46\pm 0.29\pm 0.00\pm 0.03)\cdot 10^{-2}$
& $(-3.57\pm 0.32\pm 0.00\pm 0.01)\cdot 10^{-2}$
\\
$C_{5}$ & $(9.12\pm 0.43\pm 0.03\pm 0.13)\cdot 10^{-3}$
& $(9.70\pm 0.54\pm 0.03\pm 0.09)\cdot 10^{-3}$
\\
$C_{6}$ & $(-5.28\pm 0.67\pm 0.03\pm 0.18)\cdot 10^{-2}$
& $(-5.14\pm 0.64\pm 0.03\pm 0.15)\cdot 10^{-2}$
\\
$C_{7}$ & $(0.03\pm 0.01\pm 0.19\pm 0.02)\cdot 10^{-3}$
& $(0.04\pm 0.01\pm 0.20\pm 0.02)\cdot 10^{-3}$
\\
$C_{8}$ & $(0.73\pm 0.10\pm 0.14\pm 0.07)\cdot 10^{-3}$
& $(0.69\pm 0.10\pm 0.12\pm 0.07)\cdot 10^{-3}$
\\
$C_{9}$ & $(-8.11\pm 0.08\pm 0.83\pm 0.19)\cdot 10^{-3}$
& $(-8.18\pm 0.08\pm 0.84\pm 0.19)\cdot 10^{-3}$
\\
\hline\end{tabular}
\caption[]{{\it Wilson coefficients of the effective $\Delta S=1$
Hamiltonian in $RI$ for $\lambda^*=0$ and $1$.
The first error comes from the uncertainty on $\Lambda_{QCD}$, the
second one from that on the top mass. The third error is an estimate
of the uncertainty coming from the use of different formulae
for the coefficient functions, which are equivalent at NLO order.}}
\end{center}
\protect\label{tab:cori}
\end{table}}

{\scriptsize
\begin{table}
\begin{center}
\begin{tabular}{|c|c|c|}\hline
 & Feynman RI & Landau RI \\\hline\hline
& \multicolumn{2}{c|}{$\mu=4.5$ GeV}\\\hline
$C_{1}$ & $(-1.29\pm 0.10\pm 0.00\pm 0.02)\cdot 10^{-1}$
& $(-1.58\pm 0.14\pm 0.00\pm 0.01)\cdot 10^{-1}$
\\
$C_{2}$ & $(102.92\pm 0.21\pm 0.00\pm 0.17)\cdot 10^{-2}$
& $(106.63\pm 0.64\pm 0.00\pm 0.27)\cdot 10^{-2}$
\\
$C_{3}$ & $(12.38\pm 0.97\pm 0.32\pm 0.29)\cdot 10^{-3}$
& $(1.35\pm 0.12\pm 0.03\pm 0.02)\cdot 10^{-2}$
\\
$C_{4}$ & $(-3.40\pm 0.28\pm 0.02\pm 0.03)\cdot 10^{-2}$
& $(-3.53\pm 0.30\pm 0.02\pm 0.01)\cdot 10^{-2}$
\\
$C_{5}$ & $(8.78\pm 0.42\pm 0.03\pm 0.11)\cdot 10^{-3}$
& $(9.31\pm 0.52\pm 0.03\pm 0.06)\cdot 10^{-3}$
\\
$C_{6}$ & $(-4.88\pm 0.59\pm 0.03\pm 0.16)\cdot 10^{-2}$
& $(-4.75\pm 0.56\pm 0.03\pm 0.13)\cdot 10^{-2}$
\\
$C_{7}$ & $(0.05\pm 0.01\pm 0.19\pm 0.02)\cdot 10^{-3}$
& $(0.05\pm 0.01\pm 0.20\pm 0.02)\cdot 10^{-3}$
\\
$C_{8}$ & $(0.68\pm 0.09\pm 0.12\pm 0.07)\cdot 10^{-3}$
& $(0.63\pm 0.08\pm 0.12\pm 0.07)\cdot 10^{-3}$
\\
$C_{9}$ & $(-0.95\pm 0.00\pm 0.10\pm 0.02)\cdot 10^{-2}$
& $(-0.98\pm 0.01\pm 0.10\pm 0.02)\cdot 10^{-2}$
\\
$C_{10}$ & $(1.34\pm 0.10\pm 0.17\pm 0.03)\cdot 10^{-3}$
& $(1.60\pm 0.13\pm 0.20\pm 0.03)\cdot 10^{-3}$
\\
\hline\hline
& \multicolumn{2}{c|}{$\mu=5.0$ GeV}\\\hline
$C_{1}$ & $(-11.85\pm 0.94\pm 0.00\pm 0.17)\cdot 10^{-2}$
& $(-1.46\pm 0.13\pm 0.00\pm 0.01)\cdot 10^{-1}$
\\
$C_{2}$ & $(102.50\pm 0.17\pm 0.00\pm 0.16)\cdot 10^{-2}$
& $(106.05\pm 0.57\pm 0.00\pm 0.26)\cdot 10^{-2}$
\\
$C_{3}$ & $(11.75\pm 0.89\pm 0.32\pm 0.27)\cdot 10^{-3}$
& $(1.27\pm 0.11\pm 0.03\pm 0.02)\cdot 10^{-2}$
\\
$C_{4}$ & $(-3.25\pm 0.26\pm 0.02\pm 0.02)\cdot 10^{-2}$
& $(-3.36\pm 0.28\pm 0.02\pm 0.01)\cdot 10^{-2}$
\\
$C_{5}$ & $(8.53\pm 0.43\pm 0.03\pm 0.10)\cdot 10^{-3}$
& $(9.03\pm 0.52\pm 0.03\pm 0.05)\cdot 10^{-3}$
\\
$C_{6}$ & $(-4.57\pm 0.53\pm 0.03\pm 0.14)\cdot 10^{-2}$
& $(-4.45\pm 0.50\pm 0.03\pm 0.12)\cdot 10^{-2}$
\\
$C_{7}$ & $(0.06\pm 0.01\pm 0.20\pm 0.02)\cdot 10^{-3}$
& $(0.07\pm 0.01\pm 0.20\pm 0.02)\cdot 10^{-3}$
\\
$C_{8}$ & $(0.63\pm 0.08\pm 0.12\pm 0.07)\cdot 10^{-3}$
& $(0.59\pm 0.07\pm 0.11\pm 0.07)\cdot 10^{-3}$
\\
$C_{9}$ & $(-0.94\pm 0.00\pm 0.10\pm 0.02)\cdot 10^{-2}$
& $(-0.97\pm 0.01\pm 0.10\pm 0.02)\cdot 10^{-2}$
\\
$C_{10}$ & $(1.27\pm 0.09\pm 0.16\pm 0.03)\cdot 10^{-3}$
& $(1.52\pm 0.12\pm 0.19\pm 0.03)\cdot 10^{-3}$
\\
\hline\hline
& \multicolumn{2}{c|}{$\mu=10.0$ GeV}\\\hline
$C_{1}$ & $(-5.68\pm 0.41\pm 0.00\pm 0.10)\cdot 10^{-2}$
& $(-7.78\pm 0.61\pm 0.00\pm 0.07)\cdot 10^{-2}$
\\
$C_{2}$ & $(100.19\pm 0.02\pm 0.00\pm 0.13)\cdot 10^{-2}$
& $(102.95\pm 0.23\pm 0.00\pm 0.19)\cdot 10^{-2}$
\\
$C_{3}$ & $(8.28\pm 0.51\pm 0.30\pm 0.16)\cdot 10^{-3}$
& $(8.84\pm 0.60\pm 0.32\pm 0.09)\cdot 10^{-3}$
\\
$C_{4}$ & $(-2.35\pm 0.17\pm 0.02\pm 0.01)\cdot 10^{-2}$
& $(-2.41\pm 0.18\pm 0.02\pm 0.00)\cdot 10^{-2}$
\\
$C_{5}$ & $(6.77\pm 0.39\pm 0.03\pm 0.05)\cdot 10^{-3}$
& $(7.05\pm 0.43\pm 0.04\pm 0.02)\cdot 10^{-3}$
\\
$C_{6}$ & $(-2.97\pm 0.27\pm 0.02\pm 0.08)\cdot 10^{-2}$
& $(-2.91\pm 0.26\pm 0.02\pm 0.07)\cdot 10^{-2}$
\\
$C_{7}$ & $(0.20\pm 0.00\pm 0.20\pm 0.02)\cdot 10^{-3}$
& $(0.21\pm 0.00\pm 0.21\pm 0.02)\cdot 10^{-3}$
\\
$C_{8}$ & $(4.17\pm 0.41\pm 0.77\pm 0.55)\cdot 10^{-4}$
& $(3.87\pm 0.38\pm 0.70\pm 0.55)\cdot 10^{-4}$
\\
$C_{9}$ & $(-9.11\pm 0.01\pm 0.98\pm 0.17)\cdot 10^{-3}$
& $(-0.94\pm 0.00\pm 0.10\pm 0.02)\cdot 10^{-2}$
\\
$C_{10}$ & $(7.67\pm 0.47\pm 0.99\pm 0.45)\cdot 10^{-4}$
& $(0.95\pm 0.07\pm 0.12\pm 0.05)\cdot 10^{-3}$
\\
\hline\end{tabular}
\caption[]{{\it Same as in table 4
 at different values of the scale $\mu$.}}
\end{center}
\protect\label{tab:cori2}
\end{table}}
 At the end of this section,
 we will give the practical rules to be followed, in order to obtain the
 physical amplitudes.
\par The construction of the renormalized operators appearing in (\ref{eh})
is complicated by their mixing and by the presence of the so-called
 ``effervescent" operators. To simplify the discussion, we start by considering
the $\Delta I=3/2$ ($\Delta S=2$) operator,
which, in the absence of ``effervescent"
operators, renormalizes multiplicatively.  The four-fermion $\Delta I=3/2$
operator mixes at one loop with other (``effervescent")
operators, due to
the artefacts of the regularization. This problem is common to the continuum
and
to the lattice  cases\footnote{``Effervescent" operators
appear in all  regularizations  in the presence of $\gamma_5$.
They are also present using  the procedure of ref. \cite{curci},
which avoids to define $\gamma_5$.}.
 The first step is to
 introduce  counter-terms to eliminate this artificial mixing.
This can be
achieved in many different ways. On the lattice, for example, it is obtained
by using eq. (\ref{eq:prime}). In dimensional regularizations, the elimination
of the unwanted operators
is ensured  by  minimal subtraction of the pole term:
 the string of gamma matrices,
proportional to $1/\epsilon$  at one loop, cancels
 all  ``effervescent" contributions
\cite{bur2}--\cite{noi}. There is some ambiguity in the  subtraction.
In eq. (\ref{eq:prime}), one can, for example, subtract all the terms but
$C_1^L O_1(a)$ or, in dimensional regularization, one
can subtract any further piece of $O(\epsilon^0)$ proportional to
the original four-fermion operator. The important point is that, after this
first
step, the subtracted operator $O_{{\rm sub}}$
is only proportional to the original one. Notice that, at one loop, the
coefficients
of the ``effervescent" operators are independent of the external states
and consequently they are gauge-independent, as explicit one-loop
calculations show. The independence from the external states  ensures
the possibility of   defining renormalized operators  with
definite chiral properties, by combining
 bare and  ``effervescent" operators in a gauge-invariant way.
At this point, we impose the
overall renormalization condition to the subtracted
four-quark Green function $\Gamma_{
O_{{\rm sub}}}(p)$,  by taking
$p^2=\mu^2$ for all the external legs\footnote{To
avoid unnecessary complications,
we take all  the  external momenta  equal to
$\mu^2$, i.e.  the scale that multiplies the coupling
constant in dimensional schemes.
 Of course, one can choose arbitrary external momenta, provided they
 make the Green
functions infrared-finite. In the present discussion, we assume that we
work in  the Euclidean space,
in order to make contact with the previous section.}.
Let us define
\bea \Lambda^\lambda_{
O_{{\rm sub}}}(p)&=& \hbox{Tr} \Bigl[\Gamma^\lambda_{
O_{{\rm sub}}}(p)\cdot\hat P \Bigr]\nn \\
&=&1+\frac{\alpha_s}{4 \pi}
\left(-\gamma^{(0)} \ln(p^2/\mu^2)+r^{HV,NDR}
 + (1-\lambda)r_\lambda \right), \protect\label{eq:ciu} \eea
where $\hat P$ is a suitable projector operator \cite{bur2}--\cite{noi} and
$\lambda$ denotes the gauge parameter defined in the previous section;
$\Lambda^\lambda_{
O_{{\rm sub}}}(p)$ includes the renormalization of the external quark legs.
In eq. (\ref{eq:ciu}),
$r^{HV,NDR}$ depends on the regularization in which $\Lambda^\lambda_{
O_{{\rm sub}}}(p)$
is calculated, whereas $r_\lambda$ is re\-gu\-la\-ri\-za\-tion- and
gauge-independent.  On the lattice, eq. (\ref{eq:ciu}) has a similar
expression
with $\ln( p^2/\mu^2) \to \ln (p a/\pi)^2$ and $r^{HV,NDR} \to r^{L}$.
The renormalization condition
\be \left(Z^{RI}_O(\mu, \lambda^*)\right)^{-1}
\times \Lambda^\lambda_{
O_{{\rm sub}}}(p)\vert_{p^2=\mu^2, \lambda=\lambda^*}= 1
\protect\label{eq:rc} \ee
 defines the renormalized operator
\be O^{RI}(\mu,\lambda^*)=\left(Z^{RI}_O(\mu, \lambda^*)\right)^{-1}
O_{{\rm sub}}.\ee
Then, one has
\be \Lambda^\lambda_{
O^{RI}}(p)= 1+\frac{\alpha_s}{4 \pi}
\left(-\gamma^{(0)} \ln(p^2/\mu^2)+
  (\lambda^*-\lambda)r_\lambda \right), \protect\label{eq:ciu1} \ee
where $\lambda^*$ is the gauge introduced in eq. (\ref{eq:rc}); $\lambda$ is
instead the gauge in which  the Green function, containing the insertion of
$ O^{RI}(\mu,\lambda^*)$, is computed.
\par If we call $r^{RI}$ the finite part of the matrix element of
$O^{RI}$, from eq. (\ref{eq:ciu1}) we find
\be \Delta  r^{HV,NDR}_{\lambda^*}=
 r^{RI}-r^{HV,NDR}=- r^{HV,NDR}+(\lambda^*-1) r_\lambda.
\protect\label{eq:ciu2} \ee
Equation (\ref{eq:ciu2}) has a very simple interpretation: the first term,
$- r^{HV,NDR}$,
removes the regularization dependence; the second one, $r_\lambda$,
 introduces a dependence
on $\lambda^*$; $r_\lambda$
 depends  on the external states but not
on the regularization.
The ambiguity of the first step, i.e. in the subtraction
of the pole term,  discussed above,
 is removed by the term $-r^{HV,NDR}$, which depends
on the subtraction procedure. \par
Almost all the perturbative lattice calculations have been done
in the Feynman gauge, while the Landau gauge
is the most convenient one  to implement
the non-perturbative method of ref. \cite{nonper}. For this reason,
 we give the Wilson coefficients for $\lambda^*=0,1$.
{ \scriptsize
\begin{table}
\begin{center}
\begin{tabular}{|c|c|c|c|}\hline
 & LO & NLO HV & NLO NDR\\\hline\hline
& \multicolumn{3}{c|}{$\mu=1.5$ GeV}\\\hline
$C_{1}$ & $(-4.22\pm 0.65\pm 0.00)\times 10^{-1}$
& $(-3.91\pm 0.51\pm 0.00)\times 10^{-1}$
& $(-3.80\pm 0.55\pm 0.00)\times 10^{-1}$
\\
$C_{2}$ & $(11.62\pm 0.38\pm 0.00)\times 10^{-1}$
& $(106.13\pm 0.82\pm 0.00)\times 10^{-2}$
& $(11.95\pm 0.35\pm 0.00)\times 10^{-1}$
\\
$C_{3}$ & $(1.99\pm 0.35\pm 0.00)\times 10^{-2}$
& $(2.17\pm 0.41\pm 0.00)\times 10^{-2}$
& $(2.60\pm 0.52\pm 0.00)\times 10^{-2}$
\\
$C_{4}$ & $(-4.16\pm 0.56\pm 0.02)\times 10^{-2}$
& $(-4.51\pm 0.60\pm 0.01)\times 10^{-2}$
& $(-0.63\pm 0.11\pm 0.00)\times 10^{-1}$
\\
$C_{5}$ & $(1.19\pm 0.12\pm 0.00)\times 10^{-2}$
& $(1.37\pm 0.15\pm 0.00)\times 10^{-2}$
& $(10.52\pm 0.61\pm 0.01)\times 10^{-3}$
\\
$C_{6}$ & $(-0.66\pm 0.13\pm 0.00)\times 10^{-1}$
& $(-0.63\pm 0.11\pm 0.00)\times 10^{-1}$
& $(-0.93\pm 0.21\pm 0.00)\times 10^{-1}$
\\
$C_{7}$ & $(0.16\pm 0.04\pm 0.19)\times 10^{-3}$
& $(-0.04\pm 0.00\pm 0.17)\times 10^{-3}$
& $(0.02\pm 0.06\pm 0.20)\times 10^{-3}$
\\
$C_{8}$ & $(0.63\pm 0.14\pm 0.16)\times 10^{-3}$
& $(0.97\pm 0.19\pm 0.15)\times 10^{-3}$
& $(1.06\pm 0.26\pm 0.19)\times 10^{-3}$
\\
$C_{9}$ & $(-6.77\pm 0.27\pm 0.71)\times 10^{-3}$
& $(-6.32\pm 0.37\pm 0.64)\times 10^{-3}$
& $(-7.24\pm 0.19\pm 0.73)\times 10^{-3}$
\\
\hline\hline
& \multicolumn{3}{c|}{$\mu=2$ GeV}\\\hline
$C_{1}$ & $(-3.47\pm 0.44\pm 0.00)\times 10^{-1}$
& $(-3.29\pm 0.37\pm 0.00)\times 10^{-1}$
& $(-3.13\pm 0.39\pm 0.00)\times 10^{-1}$
\\
$C_{2}$ & $(11.16\pm 0.23\pm 0.00)\times 10^{-1}$
& $(104.13\pm 0.54\pm 0.00)\times 10^{-2}$
& $(11.54\pm 0.23\pm 0.00)\times 10^{-1}$
\\
$C_{3}$ & $(1.59\pm 0.23\pm 0.00)\times 10^{-2}$
& $(1.73\pm 0.26\pm 0.00)\times 10^{-2}$
& $(2.07\pm 0.33\pm 0.00)\times 10^{-2}$
\\
$C_{4}$ & $(-3.50\pm 0.40\pm 0.01)\times 10^{-2}$
& $(-3.82\pm 0.44\pm 0.01)\times 10^{-2}$
& $(-5.19\pm 0.71\pm 0.01)\times 10^{-2}$
\\
$C_{5}$ & $(10.40\pm 0.94\pm 0.04)\times 10^{-3}$
& $(1.20\pm 0.11\pm 0.00)\times 10^{-2}$
& $(10.54\pm 0.16\pm 0.02)\times 10^{-3}$
\\
$C_{6}$ & $(-5.23\pm 0.80\pm 0.03)\times 10^{-2}$
& $(-5.08\pm 0.72\pm 0.03)\times 10^{-2}$
& $(-0.72\pm 0.13\pm 0.00)\times 10^{-1}$
\\
$C_{7}$ & $(0.18\pm 0.02\pm 0.19)\times 10^{-3}$
& $(0.01\pm 0.00\pm 0.18)\times 10^{-3}$
& $(0.01\pm 0.04\pm 0.20)\times 10^{-3}$
\\
$C_{8}$ & $(0.50\pm 0.08\pm 0.12)\times 10^{-3}$
& $(0.77\pm 0.12\pm 0.12)\times 10^{-3}$
& $(0.81\pm 0.16\pm 0.14)\times 10^{-3}$
\\
$C_{9}$ & $(-7.03\pm 0.20\pm 0.74)\times 10^{-3}$
& $(-6.71\pm 0.27\pm 0.68)\times 10^{-3}$
& $(-7.49\pm 0.15\pm 0.75)\times 10^{-3}$
\\
\hline\hline
& \multicolumn{3}{c|}{$\mu=3$ GeV}\\\hline
$C_{1}$ & $(-2.68\pm 0.28\pm 0.00)\times 10^{-1}$
& $(-2.59\pm 0.25\pm 0.00)\times 10^{-1}$
& $(-2.41\pm 0.25\pm 0.00)\times 10^{-1}$
\\
$C_{2}$ & $(10.71\pm 0.12\pm 0.00)\times 10^{-1}$
& $(101.88\pm 0.24\pm 0.00)\times 10^{-2}$
& $(11.12\pm 0.14\pm 0.00)\times 10^{-1}$
\\
$C_{3}$ & $(1.20\pm 0.14\pm 0.01)\times 10^{-2}$
& $(1.29\pm 0.16\pm 0.00)\times 10^{-2}$
& $(1.56\pm 0.19\pm 0.01)\times 10^{-2}$
\\
$C_{4}$ & $(-2.78\pm 0.26\pm 0.01)\times 10^{-2}$
& $(-3.06\pm 0.29\pm 0.01)\times 10^{-2}$
& $(-4.10\pm 0.46\pm 0.00)\times 10^{-2}$
\\
$C_{5}$ & $(8.60\pm 0.67\pm 0.04)\times 10^{-3}$
& $(9.99\pm 0.84\pm 0.04)\times 10^{-3}$
& $(9.75\pm 0.46\pm 0.03)\times 10^{-3}$
\\
$C_{6}$ & $(-3.89\pm 0.47\pm 0.03)\times 10^{-2}$
& $(-3.84\pm 0.44\pm 0.03)\times 10^{-2}$
& $(-5.33\pm 0.73\pm 0.03)\times 10^{-2}$
\\
$C_{7}$ & $(0.22\pm 0.01\pm 0.20)\times 10^{-3}$
& $(0.08\pm 0.00\pm 0.19)\times 10^{-3}$
& $(0.04\pm 0.02\pm 0.20)\times 10^{-3}$
\\
$C_{8}$ & $(3.73\pm 0.50\pm 0.92)\times 10^{-4}$
& $(5.84\pm 0.74\pm 0.91)\times 10^{-4}$
& $(0.59\pm 0.09\pm 0.10)\times 10^{-3}$
\\
$C_{9}$ & $(-7.29\pm 0.14\pm 0.77)\times 10^{-3}$
& $(-7.04\pm 0.18\pm 0.73)\times 10^{-3}$
& $(-7.66\pm 0.11\pm 0.78)\times 10^{-3}$
\\
\hline\hline
& \multicolumn{3}{c|}{$\mu=4.0$ GeV}\\\hline
$C_{1}$ & $(-2.23\pm 0.21\pm 0.00)\times 10^{-1}$
& $(-2.18\pm 0.19\pm 0.00)\times 10^{-1}$
& $(-1.99\pm 0.19\pm 0.00)\times 10^{-1}$
\\
$C_{2}$ & $(104.69\pm 0.80\pm 0.00)\times 10^{-2}$
& $(1005.88\pm 0.92\pm 0.00)\times 10^{-3}$
& $(109.00\pm 0.99\pm 0.00)\times 10^{-2}$
\\
$C_{3}$ & $(9.91\pm 0.98\pm 0.09)\times 10^{-3}$
& $(1.06\pm 0.11\pm 0.01)\times 10^{-2}$
& $(1.29\pm 0.14\pm 0.01)\times 10^{-2}$
\\
$C_{4}$ & $(-2.36\pm 0.20\pm 0.00)\times 10^{-2}$
& $(-2.61\pm 0.23\pm 0.00)\times 10^{-2}$
& $(-3.50\pm 0.35\pm 0.00)\times 10^{-2}$
\\
$C_{5}$ & $(7.48\pm 0.54\pm 0.04)\times 10^{-3}$
& $(8.75\pm 0.69\pm 0.04)\times 10^{-3}$
& $(8.97\pm 0.50\pm 0.03)\times 10^{-3}$
\\
$C_{6}$ & $(-3.18\pm 0.33\pm 0.02)\times 10^{-2}$
& $(-3.18\pm 0.32\pm 0.02)\times 10^{-2}$
& $(-4.38\pm 0.51\pm 0.02)\times 10^{-2}$
\\
$C_{7}$ & $(0.27\pm 0.01\pm 0.20)\times 10^{-3}$
& $(0.15\pm 0.00\pm 0.19)\times 10^{-3}$
& $(0.08\pm 0.01\pm 0.20)\times 10^{-3}$
\\
$C_{8}$ & $(3.10\pm 0.36\pm 0.75)\times 10^{-4}$
& $(4.87\pm 0.55\pm 0.75)\times 10^{-4}$
& $(4.76\pm 0.60\pm 0.84)\times 10^{-4}$
\\
$C_{9}$ & $(-7.44\pm 0.11\pm 0.79)\times 10^{-3}$
& $(-7.27\pm 0.15\pm 0.76)\times 10^{-3}$
& $(-7.83\pm 0.09\pm 0.80)\times 10^{-3}$
\\
\hline\end{tabular}
\caption[]{{\it Wilson coefficients for the $HV$ and $NDR$ $\overline{{\rm
MS}}$
schemes. The first error comes from the uncertainty on $\Lambda_{QCD}$, the
second one from that on the top mass.}}
\end{center}
\protect\label{tab:coms}
\end{table}}
{\scriptsize
\begin{table}
\begin{center}
\begin{tabular}{|c|c|c|c|}\hline
 & LO & NLO HV & NLO NDR\\\hline\hline
& \multicolumn{3}{c|}{$\mu=4.5$ GeV}\\\hline
$C_{1}$ & $(-2.07\pm 0.19\pm 0.00)\times 10^{-1}$
& $(-2.03\pm 0.18\pm 0.00)\times 10^{-1}$
& $(-1.84\pm 0.17\pm 0.00)\times 10^{-1}$
\\
$C_{2}$ & $(103.85\pm 0.67\pm 0.00)\times 10^{-2}$
& $(1001.15\pm 0.43\pm 0.00)\times 10^{-3}$
& $(108.20\pm 0.87\pm 0.00)\times 10^{-2}$
\\
$C_{3}$ & $(1.13\pm 0.11\pm 0.03)\times 10^{-2}$
& $(1.19\pm 0.12\pm 0.03)\times 10^{-2}$
& $(1.40\pm 0.14\pm 0.03)\times 10^{-2}$
\\
$C_{4}$ & $(-2.42\pm 0.20\pm 0.02)\times 10^{-2}$
& $(-2.67\pm 0.23\pm 0.02)\times 10^{-2}$
& $(-3.51\pm 0.34\pm 0.02)\times 10^{-2}$
\\
$C_{5}$ & $(7.06\pm 0.49\pm 0.04)\times 10^{-3}$
& $(8.36\pm 0.66\pm 0.04)\times 10^{-3}$
& $(8.75\pm 0.52\pm 0.04)\times 10^{-3}$
\\
$C_{6}$ & $(-2.94\pm 0.29\pm 0.02)\times 10^{-2}$
& $(-2.97\pm 0.29\pm 0.02)\times 10^{-2}$
& $(-4.09\pm 0.46\pm 0.02)\times 10^{-2}$
\\
$C_{7}$ & $(0.29\pm 0.01\pm 0.20)\times 10^{-3}$
& $(0.17\pm 0.00\pm 0.19)\times 10^{-3}$
& $(0.09\pm 0.01\pm 0.20)\times 10^{-3}$
\\
$C_{8}$ & $(2.88\pm 0.32\pm 0.69)\times 10^{-4}$
& $(4.53\pm 0.49\pm 0.70)\times 10^{-4}$
& $(4.39\pm 0.52\pm 0.77)\times 10^{-4}$
\\
$C_{9}$ & $(-0.96\pm 0.01\pm 0.10)\times 10^{-2}$
& $(-0.93\pm 0.00\pm 0.10)\times 10^{-2}$
& $(-0.97\pm 0.01\pm 0.10)\times 10^{-2}$
\\
$C_{10}$ & $(2.12\pm 0.19\pm 0.24)\times 10^{-3}$
& $(1.95\pm 0.17\pm 0.24)\times 10^{-3}$
& $(1.86\pm 0.16\pm 0.23)\times 10^{-3}$
\\
\hline\hline
& \multicolumn{3}{c|}{$\mu=5$ GeV}\\\hline
$C_{1}$ & $(-1.93\pm 0.17\pm 0.00)\times 10^{-1}$
& $(-1.90\pm 0.16\pm 0.00)\times 10^{-1}$
& $(-1.71\pm 0.15\pm 0.00)\times 10^{-1}$
\\
$C_{2}$ & $(103.17\pm 0.56\pm 0.00)\times 10^{-2}$
& $(9970.37\pm 0.31\pm 0.00)\times 10^{-4}$
& $(107.54\pm 0.77\pm 0.00)\times 10^{-2}$
\\
$C_{3}$ & $(10.60\pm 0.95\pm 0.34)\times 10^{-3}$
& $(1.11\pm 0.10\pm 0.03)\times 10^{-2}$
& $(1.32\pm 0.13\pm 0.03)\times 10^{-2}$
\\
$C_{4}$ & $(-2.29\pm 0.18\pm 0.02)\times 10^{-2}$
& $(-2.53\pm 0.21\pm 0.02)\times 10^{-2}$
& $(-3.32\pm 0.31\pm 0.02)\times 10^{-2}$
\\
$C_{5}$ & $(6.72\pm 0.46\pm 0.04)\times 10^{-3}$
& $(7.97\pm 0.62\pm 0.04)\times 10^{-3}$
& $(8.45\pm 0.52\pm 0.04)\times 10^{-3}$
\\
$C_{6}$ & $(-2.74\pm 0.26\pm 0.02)\times 10^{-2}$
& $(-2.78\pm 0.26\pm 0.02)\times 10^{-2}$
& $(-3.83\pm 0.41\pm 0.02)\times 10^{-2}$
\\
$C_{7}$ & $(0.31\pm 0.01\pm 0.20)\times 10^{-3}$
& $(0.20\pm 0.00\pm 0.19)\times 10^{-3}$
& $(0.11\pm 0.01\pm 0.20)\times 10^{-3}$
\\
$C_{8}$ & $(2.70\pm 0.29\pm 0.65)\times 10^{-4}$
& $(4.25\pm 0.44\pm 0.65)\times 10^{-4}$
& $(4.09\pm 0.46\pm 0.71)\times 10^{-4}$
\\
$C_{9}$ & $(-0.95\pm 0.01\pm 0.10)\times 10^{-2}$
& $(-9.23\pm 0.02\pm 1.00)\times 10^{-3}$
& $(-0.96\pm 0.01\pm 0.10)\times 10^{-2}$
\\
$C_{10}$ & $(2.00\pm 0.18\pm 0.23)\times 10^{-3}$
& $(1.87\pm 0.15\pm 0.22)\times 10^{-3}$
& $(1.77\pm 0.14\pm 0.22)\times 10^{-3}$
\\
\hline\hline
& \multicolumn{3}{c|}{$\mu=10$ GeV}\\\hline
$C_{1}$ & $(-11.76\pm 0.90\pm 0.00)\times 10^{-2}$
& $(-11.74\pm 0.88\pm 0.00)\times 10^{-2}$
& $(-9.73\pm 0.77\pm 0.00)\times 10^{-2}$
\\
$C_{2}$ & $(99.56\pm 0.11\pm 0.00)\times 10^{-2}$
& $(97.47\pm 0.18\pm 0.00)\times 10^{-2}$
& $(104.01\pm 0.34\pm 0.00)\times 10^{-2}$
\\
$C_{3}$ & $(6.86\pm 0.49\pm 0.32)\times 10^{-3}$
& $(7.12\pm 0.53\pm 0.31)\times 10^{-3}$
& $(8.74\pm 0.67\pm 0.32)\times 10^{-3}$
\\
$C_{4}$ & $(-1.52\pm 0.11\pm 0.02)\times 10^{-2}$
& $(-1.69\pm 0.13\pm 0.02)\times 10^{-2}$
& $(-2.29\pm 0.18\pm 0.02)\times 10^{-2}$
\\
$C_{5}$ & $(4.67\pm 0.30\pm 0.04)\times 10^{-3}$
& $(5.56\pm 0.40\pm 0.04)\times 10^{-3}$
& $(6.42\pm 0.41\pm 0.04)\times 10^{-3}$
\\
$C_{6}$ & $(-1.71\pm 0.14\pm 0.02)\times 10^{-2}$
& $(-1.75\pm 0.14\pm 0.02)\times 10^{-2}$
& $(-2.48\pm 0.22\pm 0.02)\times 10^{-2}$
\\
$C_{7}$ & $(0.46\pm 0.00\pm 0.20)\times 10^{-3}$
& $(0.39\pm 0.01\pm 0.20)\times 10^{-3}$
& $(0.27\pm 0.00\pm 0.20)\times 10^{-3}$
\\
$C_{8}$ & $(1.76\pm 0.15\pm 0.40)\times 10^{-4}$
& $(2.74\pm 0.23\pm 0.41)\times 10^{-4}$
& $(2.57\pm 0.23\pm 0.44)\times 10^{-4}$
\\
$C_{9}$ & $(-9.06\pm 0.04\pm 1.00)\times 10^{-3}$
& $(-8.87\pm 0.02\pm 0.98)\times 10^{-3}$
& $(-9.19\pm 0.04\pm 1.00)\times 10^{-3}$
\\
$C_{10}$ & $(1.33\pm 0.10\pm 0.15)\times 10^{-3}$
& $(1.24\pm 0.09\pm 0.15)\times 10^{-3}$
& $(1.17\pm 0.08\pm 0.15)\times 10^{-3}$
\\
\hline
\end{tabular}
\caption[]{{\it Same as in table 6  but at different scales $\mu$.}}
\end{center}
\protect\label{tab:coms2}
\end{table}
}
\par  In tables 4--7,
the  coefficients are given at several values of $\mu$,
in the $RI$ ($\lambda^*=0,1$), $NDR$ and  $HV$
renormalization schemes.
 We have extended the range of $\mu$, at
which the coefficients are computed, up to $10$ GeV, so that they can
also be used for  $B$-decays.
We present  separately
 the error coming from the uncertainty on   $\Lambda_{QCD}$,
on the value of  the top quark mass.
In computing the coefficients, several
expressions, which are equivalent at the order at which we are working, can
be used. For this reason, one may find smallish differences (of
$O(\alpha_s^2)$)
 between our
results and those presented in ref. \cite{bur3}.
The third error in tables 4 and 5
is an estimate of these $O(\alpha_s^2)$ effects, that  have been evaluated
by using for the Wilson coefficients different expressions which are
 equivalent at  the NLO.
\par We summarize the $RI$ renormalization prescription
\begin{enumerate}
\item Define a subtracted operator by removing the pole terms and
the mixing with ``effervescent" operators. Any further finite subtraction
proportional to the original operator is immaterial.
As explained in the previous section,
by working
with bare lattice operators, one only needs
 to eliminate the ``effervescent" terms.
\item Impose the renormalization condition (\ref{eq:rc}) to the subtracted
operator in a fixed gauge $\lambda^*$.
\item Take the coefficients from table 4 or
5 (when $\lambda^*=0,1$).
Notice that the non-perturbative method \cite{nonper} cannot be applied in
the Coulomb gauge.
For other gauges some more analytic work is needed.
\end{enumerate}
\par
\begin{figure}   
    \centering
    \epsfxsize=0.50\textwidth
    \leavevmode\epsffile{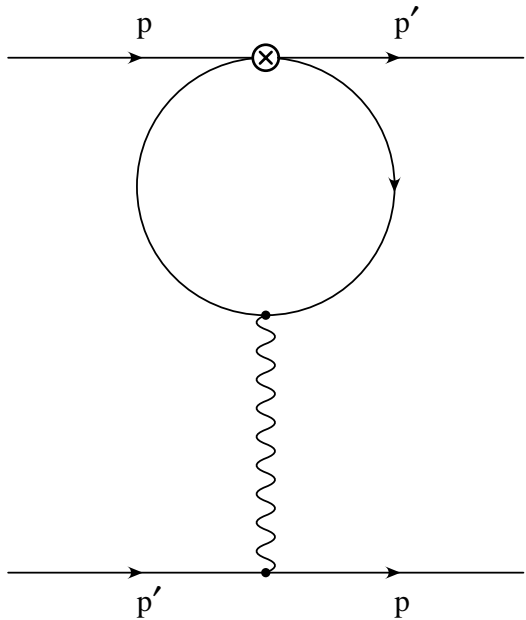}
       \caption[]{\it{Penguin diagram. We show explicitly
the external momenta used in the $RI$ renormalization
scheme.}}
    \protect\label{fig:pen}
\end{figure}

 In the more complicated case of operator mixing, i.e. mixing
between operators
belonging to the physical basis, e.g.  (\ref{epsilonprime_basis}),
 one may proceed in close analogy,
by imposing that  each renormalized  operator of the physical basis
has non-zero projection  only
onto itself.  For penguin diagrams,
 we cannot take all external legs
with  the same momentum. We then choose the convenient
configuration of momenta
given in fig. \ref{fig:pen}.
 For completeness
the matrices $\Delta  r^{HV,NDR}_{\lambda^*}$, for $\lambda^*=0,1$
are given in table 8.

\def\sln {{\ln 2}}
{ \footnotesize
\begin{table}
\begin{center}
\begin{tabular}{|c|c|c|c|c|}\hline
$(i,j)$ & $\Delta r^{HV}_{\lambda^*=0}$ & $\Delta r^{HV}_{\lambda^*=1}$
& $\Delta r^{NDR}_{\lambda^*=0}$ & $\Delta r^{NDR}_{\lambda^*=1}$\\
\hline\hline
(1,1) & $-7+4\sln$ & $-\frac{37}{6}+\frac{16}{3}
\sln$ & $-\frac{7}{3}+4
\sln$ & $-\frac{3}{2}+\frac{16}{3}\sln$\\
(1,2) & $5-12\sln$ & $\frac{13}{2}-16\sln$ &
$7-12\sln$ & $\frac{17}{2}
-16\sln$\\
(2,1) & $5-12\sln$ & $\frac{13}{2}-16\sln$ &
$7-12\sln$ & $\frac{17}{2}
-16\sln$\\
(2,2) & $-7+4\sln$ & $-\frac{37}{6}+\frac{16}{3}\sln$ & $-\frac{7}{3}+4
\sln$ & $-\frac{3}{2}+\frac{16}{3}\sln$\\
(2,3) & $-\frac{5}{27}$ & $-\frac{5}{27}$ & $-\frac{2}{27}$ & $-\frac{2}{27}$\\
(2,4) & $\frac{5}{9}$ & $\frac{5}{9}$ & $\frac{2}{9}$ & $\frac{2}{9}$\\
(2,5) & $-\frac{5}{27}$ & $-\frac{5}{27}$ & $-\frac{2}{27}$ & $-\frac{2}{27}$\\
(2,6) & $\frac{5}{9}$ & $\frac{5}{9}$ & $\frac{2}{9}$ & $\frac{2}{9}$\\
(3,3) & $-\frac{199}{27}+4\sln$ & $-\frac{353}{54}+\frac{16}{3}
\sln$ &
$-\frac{67}{27}+4\sln$ & $-\frac{89}{54}+\frac{16}{3}\sln$\\
(3,4) & $\frac{55}{9}-12\sln$ & $\frac{137}{18}-16\sln$ & $\frac{67}{9}
-12\sln$ & $\frac{161}{18}-16\sln$\\
(3,5) & $-\frac{10}{27}$ & $-\frac{10}{27}$ & $-\frac{4}{27}$ &
$-\frac{4}{27}$\\
(3,6) & $\frac{10}{9}$ & $\frac{10}{9}$ & $\frac{4}{9}$ & $\frac{4}{9}$\\
(4,3) & $5-\frac{5f}{27}-12\sln$ & $\frac{13}{2}-\frac{5f}{27}-16\sln$ &
$7-\frac{5f}{27}-12\sln$ & $\frac{17}{2}-\frac{5f}{27}-16\sln$\\
(4,4) & $-7+\frac{5f}{9}+4\sln$ & $-\frac{37}{6}+\frac{5f}{9}+\frac{16}{3}
\sln$ & $-\frac{7}{3}+\frac{5f}{9}+4\sln$ & $-\frac{3}{2}+\frac{5f}{9}
+\frac{16}{3}\sln$\\
(4,5) & $-\frac{5f}{27}$ & $-\frac{5f}{27}$ & $-\frac{5f}{27}$ &
$-\frac{5f}{27}$\\
(4,6) & $\frac{5f}{9}$ & $\frac{5f}{9}$ & $\frac{5f}{9}$ & $\frac{5f}{9}$\\
(5,5) & $-\frac{8}{3}+\frac{2}{3}\sln$ & $-\frac{7}{6}+\frac{4}{3}\sln$ &
$\frac{2}{3}+\frac{2}{3}\sln$ & $\frac{13}{6}+\frac{4}{3}\sln$\\
(5,6) & $-8-2\sln$ & $-\frac{17}{2}-4\sln$ & $-2-2\sln$ & $-\frac{5}{2}
-4\sln$\\
(6,3) & $-\frac{5f}{27}$ & $-\frac{5f}{27}$ & $-\frac{5f}{27}$ &
$-\frac{5f}{27}$ \\
(6,4) & $\frac{5f}{9}$ & $\frac{5f}{9}$ & $\frac{5f}{9}$ & $\frac{5f}{9}$ \\
(6,5) & $-2-\frac{5f}{27}-2\sln$ & $-1-\frac{5f}{27}-4\sln$ &
$2-\frac{5f}{27}-2\sln$ & $3-\frac{5f}{27}-4\sln$\\
(6,6) & $-\frac{62}{3}+\frac{5f}{9}+\frac{2}{3}\sln$ &
$-\frac{71}{3}+\frac{5f}{9}+\frac{4}{3}\sln$ &
$-\frac{34}{3}+\frac{5f}{9}+\frac{2}{3}\sln$ &
$-\frac{43}{3}+\frac{5f}{9}+\frac{4}{3}\sln$\\
(7,7) & $-\frac{8}{3}+\frac{2}{3}\sln$ & $-\frac{7}{6}+\frac{4}{3}\sln$ &
$\frac{2}{3}+\frac{2}{3}\sln$ & $\frac{13}{6}+\frac{4}{3}\sln$\\
(7,8) & $-8-2\sln$ & $-\frac{17}{2}-4\sln$ & $-2-2\sln$ & $-\frac{5}{2}
-4\sln$\\
(8,3) & $-\frac{5u}{27}+\frac{5d}{54}$ & $-\frac{5u}{27}+\frac{5d}{54}$ &
$-\frac{5u}{27}+\frac{5d}{54}$ & $-\frac{5u}{27}+\frac{5d}{54}$ \\
(8,4) & $\frac{5u}{9}-\frac{5d}{18}$ & $\frac{5u}{9}-\frac{5d}{18}$ &
$\frac{5u}{9}-\frac{5d}{18}$ & $\frac{5u}{9}-\frac{5d}{18}$ \\
(8,5) & $-\frac{5u}{27}+\frac{5d}{54}$ & $-\frac{5u}{27}+\frac{5d}{54}$ &
$-\frac{5u}{27}+\frac{5d}{54}$ & $-\frac{5u}{27}+\frac{5d}{54}$ \\
(8,6) & $\frac{5u}{9}-\frac{5d}{18}$ & $\frac{5u}{9}-\frac{5d}{18}$ &
$\frac{5u}{9}-\frac{5d}{18}$ & $\frac{5u}{9}-\frac{5d}{18}$ \\
(8,7) & $-2-2\sln$ & $-1-4\sln$ & $2-2\sln$ & $3-4\sln$\\
(8,8) & $-\frac{62}{3}+\frac{2}{3}\sln$ & $-\frac{71}{3}+\frac{4}{3}\sln$ &
$-\frac{34}{3}+\frac{2}{3}\sln$ & $-\frac{43}{3}+\frac{4}{3}\sln$\\
(9,3) & $\frac{5}{27}$ & $\frac{5}{27}$ & $\frac{2}{27}$ & $\frac{2}{27}$\\
(9,4) & $-\frac{5}{9}$ & $-\frac{5}{9}$ & $-\frac{2}{9}$ & $-\frac{2}{9}$\\
(9,5) & $\frac{5}{27}$ & $\frac{5}{27}$ & $\frac{2}{27}$ & $\frac{2}{27}$\\
(9,6) & $-\frac{5}{9}$ & $-\frac{5}{9}$ & $-\frac{2}{9}$ & $-\frac{2}{9}$\\
(9,9) & $-7+4\sln$ & $-\frac{37}{6}+\frac{16}{3}\sln$ & $-\frac{7}{3}
+4\sln$ & $-\frac{3}{2}+\frac{16}{3}\sln$\\
(9,10) & $5-12\sln$ & $\frac{13}{2}-16\sln$ & $7-12\sln$ & $\frac{17}{2}
-16\sln$\\
\hline
\end{tabular}
\caption[]{\it The matrices $\Delta r^{HV}_{\lambda^*=0,1}$
and $\Delta r^{NDR}_{\lambda^*=0,1}$
necessary to relate the Wilson coefficients in the $HV$ and $NDR$
$\overline{ MS}$ schemes
to those computed in the $RI$ scheme, in the Landau ($\lambda^*=0$) and Feynman
($\lambda^*=1$)
gauges, are given. $(i,j)$ refers to the matrix element.
The matrix elements which are zero are not reported. (Continue)}
\end{center}
\protect\label{tab:appe}
\end{table}
\addtocounter{table}{-1}
\begin{table}
\begin{center}
\begin{tabular}{|c|c|c|c|c|}\hline
$(i,j)$ & $\Delta r^{HV}_{\lambda^*=0}$ & $\Delta r^{HV}_{\lambda^*=1}$
& $\Delta r^{NDR}_{\lambda^*=0}$ & $\Delta r^{NDR}_{\lambda^*=1}$\\
\hline\hline
(10,3) & $-\frac{5u}{27}+\frac{5d}{54}$ & $-\frac{5u}{27}+\frac{5d}{54}$ &
$-\frac{5u}{27}+\frac{5d}{54}$ & $-\frac{5u}{27}+\frac{5d}{54}$\\
(10,4) & $\frac{5u}{9}-\frac{5d}{18}$ & $\frac{5u}{9}-\frac{5d}{18}$ &
$\frac{5u}{9}-\frac{5d}{18}$ & $\frac{5u}{9}-\frac{5d}{18}$\\
(10,5) & $-\frac{5u}{27}+\frac{5d}{54}$ & $-\frac{5u}{27}+\frac{5d}{54}$ &
$-\frac{5u}{27}+\frac{5d}{54}$ & $-\frac{5u}{27}+\frac{5d}{54}$\\
(10,6) & $\frac{5u}{9}-\frac{5d}{18}$ & $\frac{5u}{9}-\frac{5d}{18}$ &
$\frac{5u}{9}-\frac{5d}{18}$ & $\frac{5u}{9}-\frac{5d}{18}$\\
(10,9) & $5-12\sln$ & $\frac{13}{2}-16\sln$ & $7-12\sln$ & $\frac{17}{2}
-16\sln$\\
(10,10) & $-7+4\sln$ & $-\frac{37}{6}+\frac{16}{3}\sln$ & $-\frac{7}{3}
+4\sln$ & $-\frac{3}{2}+\frac{16}{3}\sln$\\
\hline
\end{tabular}
\caption[]{{\it (Continued) }}
\end{center}
\end{table}
}
\section{$B$-parameters of the relevant operators}
\protect\label{sec:bpar}
Since our last study of $\epse$ \cite{ciuc1}, no significant progress,
in lattice
 calculations of hadronic matrix elements, has been made. We want,
however,  to discuss a few points not clarified
in our previous papers  or  raised by other authors.
\begin{enumerate}
\item In ref. \cite{shar4}, for the
kaon $B$-parameter in $NDR$,  they  found $B_K(\mu=2$ GeV$)=0.616\pm
0.020 \pm 0.017$. Using this value and  the LO conversion factor,
 they derive the renormalization group-invariant $B_K=0.825\pm 0.027\pm 0.023$.
For consistency with the present application,
 since  corrections of $O(\alpha_s)$ were included in the
lattice calculation, we have to use
the NLO expression (\ref{eq:rgibk}) instead. In this case,
one obtains  $B_K=0.75$ as central value.
We have thus used $B_K=0.75$, assuming  an error of about
$0.15$, because  of  uncertainties coming from the
quenched approximation,
 the scale in the coupling constant, $\Lambda_{QCD}$, etc.
\item In the absence of electromagnetic corrections, it is possible to
demonstrate (non-perturbatively) that \be B^{3/2}_9(\mu) = B_K(\mu)\nn  \ee
in any regularization scheme and at any scale $\mu$ \footnote{Of course,
this is true
provided that one imposes consistent renormalization conditions.}.
The proof consists in  showing that, by assuming
$SU(3)$ flavour symmetry,  the matrix elements  $\langle
\pi^+ \pi^0 \vert Q_9 \vert K^+ \rangle_{\Delta I=3/2}$ and
$\langle
\pi^+ \pi^0\vert Q^+ \vert K^+ \rangle_{\Delta I=3/2}$
($Q^+=1/2(Q_1+Q_2)$)
  correspond to the
same expression, when written in terms of functional integrals
over quark and gluon fields. In the absence of electromagnetic
corrections and isospin breaking effects, the same two
diagrams as in  fig. \ref{fig:a2} contribute to both the decay amplitudes.
\begin{figure}   
    \centering
    \epsfxsize=1.08\textwidth
    \leavevmode\epsffile{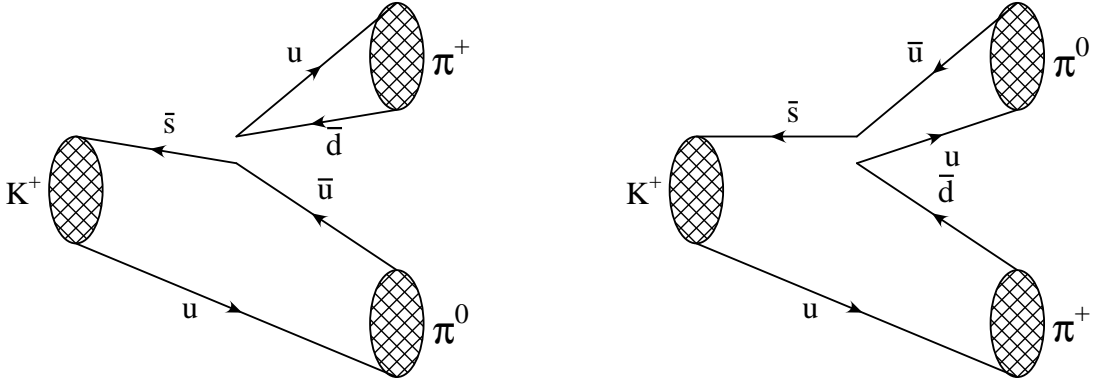}
       \caption[]{\it{Diagrams relevant for $B^{3/2}_9$ and
$(B^+)^{3/2}$  in the absence of electromagnetic corrections
and isospin breaking effects.}}
    \protect\label{fig:a2}
\end{figure}
  In the pre\-sence of $SU(3)$-brea\-king effects and of
the electromagnetic
interaction,
which has to be included for consistency with the calculation of the Wilson
coefficients, $B^{3/2}_9$ is, however, different from $(B^+)^{3/2}$.
We are not aware of any non-perturbative
estimate of  the difference of the $B$-parameters in the presence of the
electromagnetic corrections. One may
argue that the difference is small, and we have ignored it in our
analysis. \item In ref. \cite{ciuc1}, we have used  the
same value of $B^{3/2}_9$ as in the present analysis,
$B^{3/2}_9(\mu=2$ GeV$)=0.62$,
see table \ref{tab:bpar}.  This value
should be compared to $B^{3/2}_9(\mu=2$ GeV$) \sim 0.44$, which was used in
ref. \cite{burasepe}.  Our estimate of $B^{3/2}_9$ is derived from
the theoretical value of
$B_K$;
in ref. \cite{burasepe}, they evaluated  $B_9^{3/2}$ from
the experimental value of $\mbox{Re}\,A_{ 2}$  instead,
by assuming
$SU(3)$ symmetry and by neglecting electromagnetic effects.
It is  well known that, from the $\Delta I=3/2$ amplitude, and
using $SU(3)$ flavour symmetry,
one obtains a rather low value of the $B_K$-parameter, and hence
of $B_9^{3/2}=B_K$. This seems to
us  the origin of the difference.
Given the uncertainties in the extraction of
the $B$-parameters, and the electromagnetic and $SU(3)$-breaking
 effects discussed above, we think
that  the two different values of $B^{3/2}_9$
 are both  theoretically acceptable\footnote
{In ref. \cite{ciuc1}, we simply wrote $B_K=B^{3/2}_9$. This led the authors
of ref. \cite{burasepe} to believe that we had used
$B^{3/2}_9(\mu=2$ GeV$) \sim 0.8$.}.
\item In the absence of a lattice calculation,
the values of $B_{1,2}^c$  assumed in all our analyses
  have been guessed
taking into account that the corresponding matrix elements vanish
in the vacuum insertion approximation. Following
ref. \cite{reina}, we have normalized the matrix elements of
$Q_{1,2}^c$
to those of $Q_{1,2}^{(1/2)}$ by writing
$\langle Q_{1,2}^c(\mu) \rangle = B_{1,2}^c(\mu)
\langle Q_{1,2}^{(1/2)}(\mu) \rangle$.
The range of values used by us,
given in table  \ref{tab:bpar}, has been criticized  in ref. \cite{burasepe}.
Their argument is the following. By matching
at threshold ($\mu=m_c$) the effective Hamiltonian,
valid for a scale $\mu > m_c$, to the effective Hamiltonian, valid for
$\mu < m_c$,  and by assuming the positivity of
$\langle O_{\pm}^c \rangle$, they demonstrated that $B_{1,2}^c(\mu=m_c)=0$.
The values of $B_{1,2}^c(\mu>m_c)$ were then evaluated by using
the  renormalization
group evolution of the corresponding operators; they were
 found to be much smaller than
the upper values used in our case.
We believe that this demonstration is not valid
because    an important point was missed.
The effective Hamiltonian, valid for $\mu < m_c$, cannot be used to describe
the physics at threshold. This is due to  the presence of operators
of dimension higher than 6, which are usually neglected at $\mu \ll m_c$,
because their contribution decreases as $(\mu/m_c)^k$,
 with $k$ a positive integer\footnote{Indeed one
may wonder whether a region of $\mu$ where it is possible
to neglect terms of $O(\mu/m_c)$ and at the same time
to compute the Wilson coefficients in perturbation theory  really does
exist.}. Indeed, for $\mu \ll m_c$, the effect of these higher dimensional
operators enters only in the matching conditions;
at threshold, however, all the higher dimensional
operators become equally important and their contribution should be
included. This turns out to be impossible in practice.
For this reason, we believe that our guess for $B_{1,2}^c$, although with
no theoretical foundation, is an acceptable one.
\end{enumerate}
\section{Theoretical uncertainties}
\protect\label{sec:error}
In this section we briefly summarize the main sources of uncertainty present in
the determination of  $\epse$. Since many aspects of the calculation have
already
 been discussed in our previous papers on the same subject \cite{noi,ciuc1}
(see also ref. \cite{burasepe})  we only
add here a few considerations that may be useful to the reader.
\par Theoretical predictions of weak decay amplitudes rely on the
perturbative  calculation of the Wilson coefficients and on
 the non-perturbative
evaluation of the corresponding matrix elements. Even though the two aspects
are intimately connected, we discuss them separately.
\subsection{The Wilson coefficients}
\par  The main uncertainties in the evaluation of  the Wilson coefficients
are due
to  higher order corrections, which will probably remain unknown for a long
time, to the value of $\Lambda_{QCD}$, which is taken from experiments, and
to the choice
of the expansion parameter (i.e. $\alpha_s$) and  of the
renormalization scheme. The latter two  are also
related to the truncation of the perturbative expansion  and
can be reduced by increasing the scale $\mu$ at which the operators
are renormalized. It should be noticed that the results found
in $HV$ and $NDR$, for a top quark mass of $\sim 175$ GeV, differ
by about $22\%$. In principle, the differences of the Wilson
coefficients  in the two different renormalization schemes are
 compensated, at the NLO, by the
corresponding differences in  the renormalized operators, and
hence in their matrix elements.  The matrix  $\Delta r$ connecting
the two different schemes  (cf.  sec. \ref{sec:ri} and refs.
\cite{alta}--\cite{noi})  has, however,  quite large elements. This,
 combined with
the  cancellations occurring with a heavy top mass, is at the origin
of the sizeable
difference (although of
$O(\alpha_s^2)$) found  between the values of $\epse$ in $HV$ and $NDR$.
\par The larger differences observed in ref. \cite{burasepe}
between $HV$ and $NDR$
 are  due to the fact
that  the authors changed the coefficient functions
without modifying accordingly all the corresponing $B$-parameters, in
particular they kept  the values of  $B_6$ and $B_8^{3/2}$ fixed.

\subsection{Matrix elements}
\par The main
limitation to an accurate theoretical  prediction of $\epse$
does not come, however, from the calculation of the Wilson coefficients, but
it is due to
the evaluation of the operator matrix elements. We have chosen
to work in the framework of lattice $QCD$ for several reasons.
Firstly, the matching between the full theory and the effective Hamiltonian
can consistently be done at the NLO (or at any higher order).
This is not the case of other methods, like
for example the $1/N_c$-expansion. Secondly,
the calculation of the  hadronic  matrix elements of the relevant operators
is based on the fundamental
theory without extra-assumptions: there are  no
free parameters besides the quark masses
and the value of the lattice spacing, both of which are fixed by hadron
spectroscopy. Thirdly, the typical scale at which the operators are
renormalized
is of the order of the inverse lattice spacing (see sec. \ref{sec:latcon}),
 which, in current lattice
simulations, is relatively high, $1/a \sim  2$--$4$ GeV. A large
renormalization
 scale, $\mu \sim 1/a$,  reduces
the systematic error due to the presence of higher-order terms in the matching
procedure, as can be seen from the fact that the NLO corrections are smaller
at larger values of $\mu$ \cite{ciuc1,burasepe}.
Finally, the systematic errors present in lattice
calculations (finite volume effects, discretization errors, ``quenching", etc.)
can be studied and reduced in time, according to the progress realized with
 the advent of more powerful computers.
In spite of the many advantages offered by the lattice approach, the
errors in the theoretical predictions are still rather large. This is due
to a variety of reasons:
\begin{itemize}
\item
Almost all the calculations have so far been done in the so-called ``quenched"
approximation, which consists in neglecting the effects of the virtual quark
loops in the numerical simulation. This approximation is a necessity imposed
by the present limitation in computer power. The effects of the quenched
approximation on the different physical quantities is not known a priori.
In ref. \cite{ishi}, it has been shown that the value of
the kaon $B$-parameter $B_K$, obtained
in a full, ``unquenched" $QCD$ calculation with staggered fermions, does not
differ appreciably from the corresponding one obtained in
the quenched case. It should be
noticed, however, that the mass of the quarks at which the calculation of ref.
\cite{ishi}  was
performed is rather large, thus reducing the effects of the quark loops
(the quenched approximation can be obtained by  sending to infinity the mass of
the quarks in the loops).
\item Renormalized lattice operators must
correspond, in the limit of
infinite lattice cut-off, to finite chiral covariant operators, which
obey the same renormalization conditions as in the continuum.
 With few exceptions, lattice perturbation theory is
usually used to evaluate
the renormalization constants of lattice operators.
The problem of mixing with lower dimensional
operators requires, however, a non-perturbative
subtraction of all power divergences \cite{boc}, \cite{te}--\cite{martsukuba}.
Apart from this special case,
the use of perturbation theory, in the computation of
multiplicative  constants and adimensional mixing coefficients,
is well justified,
provided that the lattice spacing $a$ is sufficiently small,
i.e. $ a^{-1} \gg \Lambda_{QCD}$.
However, there is evidence of failures of lattice perturbation
theory, resulting from the use of the bare lattice
coupling constant $\alpha^{{latt}}=g_{0}^{2}/4 \pi$, as an
expansion parameter. This failure has been attributed to the presence of
``tadpole" diagrams,  which appear only in the lattice regularization and are
likely
to give large corrections in higher order perturbation theory.
Several solutions to this problem have been proposed so far
\cite{nonper,lm1},
which could hopefully reduce the systematic error induced by the poor
convergence
of lattice perturbation theory. It remains  to be shown that  these methods
 work in the case of the four-fermion operators relevant for $\epse$.
\item  In most of the lattice simulations, the $\langle \pi \vert O_i \vert
K \rangle $  matrix elements, instead of the more appropriate
$\langle \pi \pi \vert O_i \vert K \rangle $ ones, are indeed computed.
The $K \to \pi$ amplitudes are then related to the physical
$K \to \pi \pi$ ones using current algebra, at lowest order in the
chiral expansion. This certainly introduces an error, which on a single
matrix element is probably less than $30 \%$. Also in this
case, as for the isospin breaking effect mentioned before,
the systematic error is amplified by the large cancellations occurring
in the final result.
\item The crucial $B$-parameter $B_6$ has been computed so
far (together with $B_5$)  only by one group \cite{kil2,shar7},
using ``staggered" lattice fermions.
All the attempts to compute $B_{5,6}$ using different lattice
formulations of $QCD$, typically  using  ``Wilson fermions",
have failed so far. Morevover, the same techniques applied to
a similar problem, the $\Delta I=1/2$ $K \to  \pi \pi$ amplitude,
were completely unsuccessful. It would be more reassuring
to have further confirmations of the
values of $B$-parameters computed in refs.  \cite{kil2,shar7}.
\end{itemize}
\section{Conclusions}
In this paper we have discussed several theoretical issues, which
are relevant to the  prediction of  $\epse$ and many other weak decays.
To our knowledge, the discussion of the matching of the lattice operators to
the continuum effective Hamiltonian is  new.
Sections \ref{sec:latcon} and \ref{sec:ri} contain several  useful formulae
that can be used in  many different applications. With the operator
matrix elements from the lattice, one can compute
the  weak decays amplitudes,  using the  Wilson
coefficients taken from  table 4 (5) or
6 (7),
with almost no  extra effort. The $RI$ renormalization scheme discussed in
this paper will become very useful if
the non-perturbative method proposed in ref. \cite{nonper}
will succeed in the case of  four-fermion operators.
\par We have also presented an upgraded theoretical
analysis of $\epse$ in two different renormalization
schemes, $NDR$ and $HV$.
Let us summarize the main results of the present
study. Using the information coming from $\ep$ and $x_d$, and
taking $m_t=(174\pm 17)$ GeV and $f_B B^{1/2}= (200 \pm 40 )$ MeV,
we obtain constraints on the angles of the unitary triangle
\bea  \cos \delta= 0.47 \pm 0.32 \, , \nn \\
\sin 2 \beta=0.65 \pm 0.12 \, . \eea
Moreover, for $\cos \delta >0$, we predict
\be x_d=0.75 \pm 0.23 \, . \nn \ee
Finally, we have obtained
\be \epse= (3.1 \pm 2.5 ) \times 10^{-4} \nn \ee
as our best estimate. Isospin-breaking effects have been found
quite important, because of the large value  of $m_t$, and we believe
that they deserve further theoretical investigation.

\section*{Acknowledgements}
 The partial
support of  M.U.R.S.T. and of the EC contract CHRX-CT93-0132 is acknowledged.

\end{document}